\documentclass[titlepage,12pt]{article}

\usepackage{textcomp}
\usepackage{chetnew}
\usepackage{amssymb,amsmath,color,graphics,amscd,epsf,indentfirst,amsfonts}
\usepackage[utf8]{inputenc}
\usepackage{mathtools}
\usepackage{enumerate}
\usepackage{epsfig}
\usepackage{slashed}
\usepackage{tikz}

\def\blfootnote{\xdef\@thefnmark{}\@footnotetext}

\long\def\symbolfootnote[#1]#2{\begingroup%
\def\thefootnote{\fnsymbol{footnote}}\footnote[#1]{#2}\endgroup}

\makeatletter
\renewcommand{\@dotsep}{4.5}
\makeatother

\def\be{\begin{equation}}
\def\ee{\end{equation}}

\makeatletter
\def\@seccntformat#1{\csname the#1\endcsname.\quad}
\makeatother

\def\clock{{\count0=\time
           \divide\count0 60
           \ifnum\count0<10 0\fi\the\count0
           \multiply\count0 -60 \advance\count0 \time
           :\ifnum\count0<10 0\fi \the\count0
         }}
\newcommand{\timestamp}{{\small\vbox{\hbox{\tt\jobname.tex}
\hbox{\the\day/\the\month/\the\year, \clock}}}}

\def\AA{{\cal A}}
\def\BB{{\cal B}}
\def\CC{{\cal C}}
\def\DD{{\cal D}}

\def\HH{{\cal H}}

\def\JJ{{\cal J}}

\def\MM{{\cal M}}
\def\NN{{\cal N}}
\def\OO{{\cal O}}

\def\QQ{{\cal Q}}
\def\RR{{\cal R}}
\def\SS{{\cal S}}

\def\VV{{\cal V}}

\def\IR{{\mathbb R}}

\def\IZ{{\mathbb Z}}
\def\IC{{\mathbb C}}

\def\sint{{\int_{S^3}}}

\def\Tr{{\rm {Tr}}}

\def\d{{\partial}}
\def\p{{\partial}}
\def\time{{\boldsymbol{\tau}}}
\def\tt{{$tt^*$}}

\def\beq{\begin{equation}}
\def\eeq{\end{equation}}
\newcommand{\bea}{\begin{eqnarray}}
\newcommand{\eea}{\end{eqnarray}}
\def\bal{\begin{align}\begin{split}}
\newcommand{\eal}{\end{split}\end{align}}

\numberwithin{equation}{section}

\preprint{DCPT-18/23 \\ \vspace{1cm}}

\title{Geometry of Higgs-branch\\ \vspace{0.3cm} superconformal primary bundles}

\author{Vasilis Niarchos}

\affiliation{Department of Mathematical Sciences and Center for Particle Theory\\ Durham University, Durham DH1 3LE, UK
\vspace{1cm}
}

\abstract{It is known that the two- and three-point functions of Higgs-branch superconformal primaries in 4d $\NN=2$ superconformal field theories obey non-renormalization theorems on $\NN=2$ superconformal manifolds. In this paper we prove a stronger statement, that the bundles of Higgs-branch superconformal primaries over $\NN=2$ superconformal manifolds are endowed with a flat connection, or equivalently that Higgs-branch superconformal primaries have vanishing Berry phases under $\NN=2$ exactly marginal deformations. This statement fits well with the proposed correspondence between the rigid structures of 2d chiral algebras and the sector of Schur operators in 4d $\NN=2$ theories. We also discuss the general interplay between non-renormalization theorems and the curvature of bundles of protected operators and provide a new simpler proof of the vanishing curvature of 1/2-BPS operators in 4d $\NN=4$ SYM theory that does not require the use of the 4d $tt^*$ equations.}

\date{\today}

\begin{document}

\maketitle

\hypersetup{pageanchor=true}

\setcounter{tocdepth}{2}

\toc

\section{Introduction} 
\label{setup}

\subsection{Background and motivation}

In this paper we focus on four-dimensional superconformal quantum field theories (SCFTs) that possess non-trivial conformal manifolds $\MM$. By definition, each point of the space $\MM$ is a SCFT. One can move from one point of $\MM$ to another with a continuous exactly marginal deformation. As a result, $\MM$ is parametrized by the set of exactly marginal couplings that describe these deformations. The theories of interest in this paper possess exactly marginal couplings that preserve $\NN=2$ supersymmetry (i.e.\ eight Poincar\'e supercharges).

At each point of $\MM$ a SCFT comes equipped with a set of local operators, or via the operator-state correspondence, with a Hilbert space of states in radial quantization. Globally on the conformal manifold, the vector space of operators or states defines a vector bundle with base $\MM$, which is graded by the scaling dimension and other global charges of the theory. This vector bundle is endowed with a connection $\nabla$ that allows the proper comparison of data between the SCFTs at different values of the exactly marginal couplings. 

Natural notions of connection have been discussed in conformal perturbation theory, \cite{Kutasov:1988xb,Ranganathan:1993vj}. Physically, the connection is born out of the regularization of ultraviolet divergences. The naive derivative of an $n$-point function with respect to an exactly marginal coupling $\lambda^\mu$ turns into a covariant derivative on the conformal manifold {\it after regularization of the UV divergences}
\beq
\label{setaa}
\nabla_\mu \left \langle \phi_{i_1}(x_1) \cdots \phi_{i_n}(x_n) \right\rangle = \left \langle \int d^4x\, \OO_\mu (x) \phi_{i_1}(x_1) \cdots \phi_{i_n}(x_n) \right \rangle_{renormalized}
~.
\eeq
$\OO_\mu$ is the exactly marginal operator corresponding to the coupling $\lambda^\mu$. The connection $\nabla$ can be used to formulate covariant statements about the coupling constant dependence of local correlation functions on the superconformal manifold $\MM$.

Alternatively, a natural connection arises in radial quantization as the Pancharatnam-Berry connection \cite{pancharatnam,Berry:1984jv,Simon:1983mh} on the vector bundle of Hilbert spaces. This connection, which has a specific counterpart in conformal perturbation theory \cite{Baggio:2017aww}, will play a key r\^ole in the discussion of this paper. We will review the specifics of the Berry connection that we need in section \ref{berry}. Because of the operator-state correspondence, we will use interchangeably the language of operators and states.

The connection that we will employ is, by construction, metric-compatible. The notion of a metric on the vector bundle of operators arises naturally from the 2-point functions of the CFT. More explicitly, the 2-point function of two conformal primary operators $\phi_I$, $\phi_J$ with the same scaling dimension $\Delta$ is 
\beq
\label{setab}
\langle \phi_I(x) \phi_J (0) \rangle = \frac{g_{I J}}{|x|^{2\Delta}}
\eeq 
and metric-compatibility means $\nabla_\mu g_{IJ}=0$. This is a covariant derivative with respect to the exactly marginal coupling constants $\lambda^\mu$ (not to be confused with covariant derivatives in spacetime). The scaling dimension $\Delta$ of operators in short multiplets is, at least locally on the conformal manifold, independent of the couplings $\lambda^\mu$.

Three- and higher-point functions are more interesting as they capture dynamical information of the theory (and not just data related to the definition of the operators). The 3-point functions of three conformal primary operators in a CFT, say $\phi_I,\phi_J,\phi_K$, is fixed by the spacetime-independent constants $C_{IJK}$
\beq
\label{setac}
\langle \phi_I (x_1) \phi_J (x_2) \phi_K (x_3) \rangle = \frac{C_{IJK}}{|x_{12}|^{\Delta_I+\Delta_J-\Delta_K} |x_{13}|^{\Delta_I+\Delta_K-\Delta_J} |x_{23}|^{\Delta_J+\Delta_K-\Delta_I} } 
~.
\eeq
On a superconformal manifold the coefficients $C_{IJK}$ are in general non-trivial functions of the exactly marginal coupling constants. There are, however, special cases of protected operators whose 3-point functions are covariantly constant, namely they have the property 
\beq
\label{setad}
\nabla_\mu C_{IJK}=0
~.
\eeq
This property is what we usually call a {\it non-renormalization theorem for 3-point functions}. 

It is useful to stress the following aspects of the covariant formulation of such theorems:
\begin{itemize}

\item[$(a)$] When the connection $\nabla_\mu = \p_\mu + A_\mu$ is not flat, i.e.\ the curvature $F_{\mu\nu}=[\nabla_\mu,\nabla_\nu]\neq 0$, operators with the same scaling dimension undergo non-trivial mixing under parallel transport on the conformal manifold. This mixing enters, for example, in computations of correlation functions in conformal perturbation theory. However, if a quantity like the 3-point function coefficients in the non-renormalization theorem \eqref{setad} is covariantly constant, it is guaranteed that we can find a non-holonomic coupling-constant-dependent basis (like the vierbein basis for tangent vectors on a curved manifold) that undoes the mixing and makes $C_{IJK}$ independent of the exactly marginal coupling constants (this is what we typically have in mind when we say non-renormalization).  

\item[$(b)$] When the connection $\nabla$ is flat, there is no mixing under parallel transport. Then, we can choose a coupling constant-independent basis where $A_\mu=0$ and \eqref{setad} becomes simply $\p_\mu C_{IJK}=0$. Clearly, a non-renormalization theorem combined with a flat connection is a stronger statement.

\end{itemize}

A non-renormalization theorem in a protected subsector does not necessarily imply that the connection in this subsector is flat. A well-known example of a protected sector with non-vanishing curvature is the sector of 1/2-BPS chiral primary operators in 2d $\NN=(4,4)$ superconformal manifolds \cite{Gaberdiel:2007vu,Dabholkar:2007ey,Pakman:2007hn,Taylor:2007hs,deBoer:2008ss,Baggio:2012rr}. We will review some of the pertinent details of this example in subsection \ref{nonrenormImplications}. Nevertheless, non-renormalization does not come without any implications to the curvature. In subsection \ref{nonrenormImplications} we argue quite generally that the existence of a non-renormalization theorem for 3-point functions implies the integrability condition 
\beq
\label{setae}
[\nabla_\mu, \nabla_{\nu} ] C_{KL}^M =0 ~~\Leftrightarrow~~
({F}_{\mu \nu})_K^S \, C_{SL}^M + ({F}_{\mu \nu})_L^S \, C_{KS}^M 
- ({F}_{\mu \nu})_S^M \, C_{KL}^S
=0
~,
\eeq
which can lead to non-trivial relations between the curvature and 3-point function coefficients at different scaling dimensions if the curvature is non-vanishing. A similar relation is satisfied with arbitrary covariant derivatives of the curvature. In the example of chiral primary operators in 2d $\NN=(4,4)$ superconformal manifolds the covariant derivative of the curvature vanishes, hence \eqref{setae} is the only relation of this type that follows from \eqref{setad}.

Another well studied example of protected operators, whose 2- and 3-point functions do not renormalize, is the example of 1/2-BPS operators in 4d $\NN=4$ SYM theory. Initial indications of non-renormalization in this case based on the AdS/CFT correspondence appeared in \cite{Lee:1998bxa,DHoker:1998vkc,DHoker:1999jke}. More concrete arguments in favor of non-renormalization were given in \cite{Intriligator:1998ig,Intriligator:1999ff,Eden:1999gh,Petkou:1999fv,Howe:1999hz,Heslop:2001gp} and later with more elementary methods in \cite{Basu:2004nt,Baggio:2012rr}. In this case the connection on the $\NN=4$ SYM conformal manifold is flat \cite{Papadodimas:2009eu,Baggio:2014ioa}. The proof of this statement reveals interesting structures in the chiral ring of the 1/2-BPS operators and demonstrates why studies of the geometry of the bundles of operators on conformal manifolds can be useful even in cases with constrained dynamics characterized by non-renormalization theorems. The vanishing curvature was deduced in \cite{Papadodimas:2009eu,Baggio:2014ioa} by combining three elements: the $tt^*$ equations, the non-renormalization of 3-point functions, and a non-trivial identity between 2- and 3-point functions at tree-level. Let us quickly recall the main points of this argument. 

Ref.\ \cite{Papadodimas:2009eu} originally determined the curvature of 1/2-BPS chiral rings in 4d $\NN=2$ and $\NN=4$ theories in conformal perturbation theory and found it to be given solely in terms of the 2- and 3-point functions in the chiral ring. This result was re-derived in quantum mechanics from standard formulae of Berry curvature in \cite{Baggio:2017aww}. The resulting expressions are known as the $tt^*$ equations (because of their resemblance to analogous equations derived in the original work on topological-antitopological fusion in 2d $\NN=(2,2)$ topologically twisted QFTs by Cecotti and Vafa \cite{Cecotti:1991me}). In $\NN=4$ SYM theory that has a complex one-dimensional conformal manifold $\MM$ parametrized by the complexified Yang-Mills coupling $\tau = \frac{\theta}{2\pi} + \frac{4\pi i}{g_{YM}^2}$ the $tt^*$ equations read
\begin{subequations}
\beq
\label{setaf}
\left( {F}_{\tau\tau} \right)_K^L = [\nabla_\tau, \nabla_\tau]_K^L = 0~, ~~
\left( {F}_{\bar \tau \bar \tau} \right)_K^L = [\nabla_{\bar \tau}, \nabla_{\bar \tau}]_K^L = 0~, ~~
\eeq
\beq
\label{setag}
\left( {F}_{\tau \bar \tau} \right)_K^L = [\nabla_\tau, \nabla_{\bar \tau}]_K^L = - [C_\tau , \bar C_{\bar \tau}]_K^L + g_{\tau\bar \tau} \delta_K^L \left ( 1+ \frac{r}{4c} \right)
~.
\eeq
\end{subequations}
In the last equation $r$ is the common $U(1)_r$ charge of the chiral primaries $\phi_K$, $\phi_L$ and $c$ is the central charge of the SCFT, which is a constant on $\MM$. $g_{\tau\bar \tau}$ is up to an overall constant coefficient the Zamolodchikov metric on $\MM$. The first couple of equations, \eqref{setaf}, are simply the statement that the 1/2-BPS chiral primary bundles are holomorphic. This statement is trivial by anti-symmetry in $\NN=4$ SYM theory (but is less trivial in 4d $\NN=2$ theories with higher-dimensional conformal manifolds). In $\NN=4$ SYM theory even $F_{\tau\bar \tau}$ vanishes,\footnote{In general 4d $\NN=2$ SCFTs the holomorphic-antiholomorphic components of the curvature, $F_{i\bar j}$, of the corresponding Coulomb-branch chiral primaries are non-trivial functions of the exactly marginal couplings. They can be determined analytically via the \tt\ equations using supersymmetric localization techniques \cite{Gerchkovitz:2016gxx}.} although this is not automatically evident in \eqref{setag}. Non-renormalization can be used to derive this statement from the $tt^*$ equations in the following way.

The non-renormalization of 3-point functions in the 1/2-BPS chiral ring of $\NN=4$ SYM theory states that
\beq
\label{setai}
\nabla_{\tau} C_{IJK} = 0
~, ~~
\nabla_{\bar \tau} C_{IJK} =0
~,
\eeq
where the $I,J,K$ indices refer to 1/2-BPS chiral primary operators. Then, acting with a covariant derivative on \eqref{setag} we obtain, after implementing \eqref{setai}, the equations $\nabla_\tau F_{\tau\bar \tau}= \nabla_{\bar \tau} F_{\tau\bar \tau}=0$, which establish that the curvature is covariantly constant \cite{Papadodimas:2009eu}. By showing that the RHS of eq.\ \eqref{setag} is zero at tree-level, \cite{Baggio:2014ioa}, one proves that the curvature of 1/2-BPS chiral primary bundles vanishes in $\NN=4$ SYM theory.

In appendix \ref{half} we will present a much faster way to show that the curvature of 1/2-BPS chiral primary bundles vanishes by using some of the special properties of $\NN=4$ SYM theory. This alternative proof does not go though the $tt^*$ equations \eqref{setag}, the non-renormalization theorem, or the tree-level identities of \cite{Baggio:2014ioa} (see \eqref{setaj} below). Having said this, it is instructive to turn part of the above argument around.

Assume we know the $tt^*$ equations and that the curvature vanishes in the chiral ring. Then, from the RHS of eq.\ \eqref{setag} we deduce the following equation that relates 2- and 3-point functions in the 1/2-BPS chiral ring of $\NN=4$ SYM theory
\beq
\label{setaj}
\left[ C_\tau, C_{\bar \tau} \right ]_K^L = g_{\tau \bar \tau}\, \delta_K^L \left( 1+ \frac{r}{4c} \right) 
~,\eeq
or even more explicitly
\beq
\label{setak}
C_{\tau K}^P g_{P\bar Q} C^{*\bar Q}_{\bar \tau \bar R} g^{\bar R L} - g_{K\bar N} C^{*\bar N}_{\bar \tau \bar U} g^{\bar U V} C_{\tau V}^L
= g_{\tau \bar \tau}\, \delta_K^L \left( 1+ \frac{r}{4c} \right) 
~.
\eeq
In Ref.\ \cite{Baggio:2014ioa} \eqref{setaj} was proved independently at tree-level and then was argued to hold at all values of the coupling because of the non-renormalization theorem. Instead, \eqref{setaj} is deduced here from the vanishing curvature (i.e.\ the vanishing of the LHS of eq.\ \eqref{setag}). Eq.\ \eqref{setaj} is a non-trivial identity between correlation functions. Although at the end the 2- and 3-point functions in \eqref{setak} can be determined by conceptually trivial Wick contraction computations in free theory, these contractions can be combinatorially very complicated (see \cite{Corley:2001zk} for related discussions).\footnote{One can also derive the general 2- and 3-point functions of 1/2-BPS chiral primaries in $\NN=4$ SYM theory from suitable derivatives of the $S^4$ partition function using supersymmetric localization \cite{Gerchkovitz:2016gxx}.} The identity \eqref{setaj} streamlines these computations and in some cases, e.g.\ in $SU(2)$ $\NN=4$ SYM theory, it gives the full expression of 2- and 3-point functions of chiral primaries at arbitrary scaling dimension in a few lines. In fact, as was pointed out in \cite{Baggio:2015vxa} the underlying structure behind the equations \eqref{setaj} is a Heisenberg algebra. We refer the reader to \cite{Baggio:2015vxa} for further details. 

The above example demonstrates how a simple geometric statement on the curvature of a subbundle on the conformal manifold translates to non-trivial relations between correlation functions. This is one of the reasons why we want to study such geometric data more generally on superconformal manifolds, even in cases where non-renormalization properties put a tight constraint on the dynamics.

\subsection{Non-renormalization theorems in $\NN=2$ SCFTs}

Other well known protected sectors whose 2- and 3-point functions do not renormalize are the Schur operators in 4d $\NN=2$ SCFTs. For the reasons outlined above, we would like to know the specific features of the corresponding operator bundles. In section \ref{n2} we will determine the curvature of Higgs-branch superconformal primaries on 4d $\NN=2$ superconformal manifolds. As we review in a moment the Higgs-branch superconformal primaries are a special case of Schur operators. They are the only operators in the Schur class that are superconformal primaries.

In the rest of this subsection we summarize some of the defining features of the Schur operators.

\paragraph{$\NN=2$ SCFTs: Schur operators.} 
The $\NN=2$ superconformal algebra in four dimensions has an $SU(2)_R \times U(1)_r$ symmetry. The superconformal primary operators belong in representations of $SU(2)_R$ and are labelled by two half-integers $(R,m)$, where $R\in \frac{1}{2} \IZ_+$ and $m=-R,\ldots,R$. They are also labeled by the $U(1)_r$ charge $r$. 

The complete list of short multiplets of the $\NN=2$ superconformal algebra can be found in \cite{Dolan:2002zh,Cordova:2016emh}. The list includes the Schur operators, which obey the shortening conditions
\beq
\label{setba}
\Delta = 2 R + \frac{1}{2} (j+\bar j)~, ~~ r = j - \bar j
~.
\eeq   
$j$, $\bar j$ are respectively the spins of the two spacetime $SU(2)$s in the Lorentz group $SO(4)\simeq SU(2)_L \times SU(2)_R$. Following the conventions of \cite{Cordova:2016emh} both $j,\bar j$ are positive {\it integers} here. The Schur operators are 1/2-BPS and belong to the superconformal multiplets
\bea
\label{setbb}
\begin{array}{c}
B_1 \bar B_1[j=0;\bar j=0]_{\Delta=2R}^{(R;r=0)} ~, ~~ 
B_1 \bar A_1[j=0;\bar j\geq 1]_{\Delta=1+2R+\frac{1}{2}\bar j}^{(R;r=-2-\bar j)} ~, ~~
\\ \vspace{-0.3cm}\\
\bar B_1 A_1[j\geq 1;\bar j=0]_{\Delta=1+2R+\frac{1}{2} j}^{(R;r=2+ j)} ~, ~~ 
A_1 \bar A_1[j\geq 1;\bar j\geq 1]_{\Delta=2+2R+\frac{1}{2}(j+\bar j)}^{(R;r=j-\bar j)}
\end{array}
\eea
in the notation of \cite{Cordova:2016emh}.\footnote{In the notation of \cite{Dolan:2002zh} the corresponding superconformal multiplets are $\hat \BB_R$, $\DD_{R(0,\frac{1}{2}\bar j)}$, $\bar \DD_{R(\frac{1}{2} j,0)}$, $\hat \CC_{R(\frac{1}{2}j,\frac{1}{2} \bar j)}$.} The Schur operators in $B_1\bar B_1$ are the Higgs-branch superconformal primaries. Their vacuum expectation value parametrizes the Higgs branch vacua of the theory, where the $U(1)_r$ is unbroken but the $SU(2)_R$ R-symmetry is spontaneously broken.\footnote{The $\NN=2$ SCFTs also possess another 1/2-BPS sector, the sector of the Coulomb-branch chiral primaries, which was mentioned in passing in footnote 1. These are scalar superconformal primaries obeying the shortening condition $\Delta = \frac{|r|}{2}$. They have vanishing $SU(2)_R$ charge, $R=0$, and their vacuum expectation value parametrizes the vacua of the Coulomb branch of the theory.} These operators are superconformal primaries and will play a protagonistic r\^ole in the computations of section \ref{n2} below. We will not consider extensively the rest of the Schur operators, which are conformal primaries but superconformal descendants. We anticipate that the arguments of section \ref{n2} can be suitably modified to include the more general Schur operators, but we will not undertake this task in the present paper.

Repeating the logic in \cite{Baggio:2012rr} the authors of \cite{Beem:2013sza} argued that in a general 4d $\NN=2$ SCFT with exactly marginal couplings $(\lambda^i, \bar \lambda^i)$ $(i=1,\ldots,\dim_{\IC}\MM)$ the 3-point functions $C_{IJK}$ of any triplet of Schur operators (including the Higgs-branch superconformal primaries) satisfy the non-renormalization conditions
\beq
\label{setbc}
\nabla_{\lambda^i} C_{IJK} = 0~,~~ \nabla_{\bar \lambda^i} C_{IJK}=0
~.
\eeq

We remind the reader that the Schur operators are in the cohomology of a certain nilpotent supercharge, which was constructed in \cite{Beem:2013sza}. The cohomological construction in \cite{Beem:2013sza} leads to a direct relation with a two-dimensional chiral algebra that can be used to compute correlation functions of the Schur operators. The non-renormalization theorem \eqref{setbc} plays a central r\^ole in this 2d/4d correspondence since the 2d chiral algebras are rigid structures, which are independent of the exactly marginal couplings of the corresponding 4d theories. In this context, it is of interest to know the curvature of the Schur bundles, which determines whether the $C_{IJK}$ are just covariantly constant or (as a stronger statement) constant in a coupling-constant-independent basis.

\subsection{Outline of the paper}

There are two equivalent ways to evaluate curvatures on a conformal manifold. One approach, based on conformal perturbation theory (see \cite{Papadodimas:2009eu} and the original references therein), is naturally formulated in the language of local operators. Another approach, which is based on the operator-state correspondence, formulates the connection on the conformal manifold as a Berry connection. In this paper, we will adopt the second approach following a prescription previously developed and successfully tested in \cite{Baggio:2017aww}. The necessary elements of this computation are set up in section \ref{berry}. 

As we hinted in the introduction there can be an interesting interplay between the curvatures of protected operators/states and non-renormalization theorems. In subsection \ref{nonrenormImplications} we discuss a simple integrability condition on the curvature that follows from the non-renormalization of 3-point functions and we exhibit how it works in the non-trivial case of 1/2-BPS operators in 2d $\NN=(4,4)$ SCFTs.

The main computation of the paper is presented in section \ref{n2}. We compute the Berry curvature of Higgs-branch superconformal primary bundles in $\NN=2$ SCFTs and show that it vanishes. Unlike the case of 1/2-BPS operators in $\NN=4$ SYM theory, here there are no indications that the vanishing curvature relies on a set of non-trivial identities between tree-level 2- and 3-point functions. 

In section \ref{n4} we discuss a few interesting open problems highlighting the case of 1/4-BPS operators in $\NN=4$ SYM theory. 

A summary of conventions and useful technical details is relegated to two appendices at the end of the paper. In a third appendix, \ref{half}, we present a new simpler proof of the vanishing curvature of 1/2-BPS chiral primaries that does not involve the $tt^*$ equations or the non-renormalization theorem.

\section{Berry phases and non-renormalization theorems}
\label{berry}

In the introduction we emphasized the fact that the most appropriate way to think about CFT operators on a conformal manifold is in terms of sections of vector bundles. Conformal perturbation theory naturally induces a connection $\nabla$ on these vector bundles, which encodes the regularization scheme of correlation functions with insertions of spacetime-integrated exactly marginal operators.

Another route to a connection $\nabla$ on conformal manifolds is through the operator-state correspondence. In the language of states we consider vector bundles of Hilbert spaces over $\MM$. In the quantum mechanics that arises when one places the CFT on $\IR \times S^3$ one can define the Pancharatnam-Berry connection \cite{pancharatnam,Berry:1984jv}. This connection encodes the geometric phase that quantum states pick up under an adiabatic cyclic variation of the parameters in the conformal Hamiltonian. Ref.\ \cite{Baggio:2017aww} showed that the curvature of the Berry connection associated to the variation of the exactly marginal couplings is identical to the curvature of the connection that was used in \cite{Papadodimas:2009eu} in conformal perturbation theory. In the rest of the paper we will adopt the Hamiltonian approach in radial quantization and will work primarily with the corresponding Berry connection. In the next subsection we summarize the relevant notation. 

\subsection{Berry phases in radial quantization}
\label{radial}

Quite generally consider a quantum mechanics system with Hamiltonian $H$ depending on a set of parameters $\lambda^\mu$. In this paper the quantum mechanics of interest arises from a CFT in radial quantization on $\IR \times S^3$. $H$ is a suitable transformation of the dilatation operator called the conformal Hamiltonian and $\lambda^\mu$ are exactly marginal couplings. In an arbitrary subspace of the Hilbert space of states with energy $(={\rm scaling~ dimension})$ $E_n(\lambda)$ and degeneracy $d_n$ let us choose a basis of degenerate eigenstates $|n,a\rangle_\lambda$, where $a=1,\ldots, d_n$. We assume that, at least locally on the conformal manifold, there are no level crossings so that $d_n$ is independent of $\lambda$. The Berry connection and curvature in this subspace are respectively \cite{Berry:1984jv}
\beq
\label{berryaa}
\left( {A}_\mu^{(n)} \right)_{ab} = \langle n,b |\d_\mu | n, a\rangle
~,
\eeq
\beq
\label{berryab}
\left( {F}_{\mu\nu}^{(n)} \right)_a^{~b} = \d_\mu \left( {A}_\nu^{(n)} \right)_a^{~b} - \d_\nu \left( {A}_\mu^{(n)} \right)_a^{~b} - \left[ {A}_\mu^{(n)}, {A}_\nu^{(n)} \right]_a^{~b}
~.
\eeq
We raise and lower the Hilbert space indices $a,b$ with the 2-point functions
\beq
\label{berryac}
g_{(n)ab}(\lambda) = \, _\lambda \hspace{-0.02cm} \langle n,a | n, b \rangle_\lambda
~.
\eeq
A standard formula in quantum mechanics expresses the Berry curvature as an infinite sum over all intermediate states with energies different from the energy of the external states
\beq
\label{berryad}
\left( {F}_{\mu\nu}^{(n)}\right)_{ab} = \sum_{m\neq n} \sum_{c,d} \frac{1}{(E_n-E_m)^2} \langle n,b | \p_\mu H | m,c\rangle g_{(n)}^{cd} \langle m,d | \p_\nu H | n,a\rangle - (\mu \leftrightarrow \nu)
~.
\eeq

The primary goal of this paper is to evaluate the expression \eqref{berryad} in $\NN=2$ SCFTs with 1/2-BPS Higgs-branch superconformal primaries as external states. For Coulomb branch chiral primaries in $\NN=2$ SCFTs and 1/2-BPS chiral primaries in $\NN=4$ SYM theories this computation was performed in \cite{Baggio:2017aww}. We revisit the computation of 1/2-BPS chiral primaries in $\NN=4$ SYM theory in appendix \ref{half}.

\subsection{Berry phases of R-symmetry descendants}
\label{gendescend}

The superconformal primaries, whose Berry phase we will be computing in this paper, belong in some representation of the R-symmetry group of the theory. In $\NN=2$ theories this group is $SU(2)_R \times U(1)_r$. In what follows we will focus on irreducible representations of the R-symmetry group. A convenient way to construct irreducible representations is to act on a highest weight state with lowering operators in the Chevalley basis. Two different states in the same irreducible representation are characterized by different weights and cannot mix under adiabatic cycles in parameter space. Hence, they cannot exhibit non-trivial non-abelian Berry mixing. 

Moreover, since different superconformal primaries in an irreducible representation of the R-symmetry group are related by a global symmetry, it is expected that all the descendants of a highest weight state have the same Berry phase. As a check, we will verify this statement by explicit computation in subsection \ref{n2descendants}.

\subsection{Implications of covariantly constant 3-point functions}
\label{nonrenormImplications}

Let us consider a vector bundle $\VV$ of operators (or states) on a conformal manifold $\MM$. In this subsection, the sections of this bundle are protected operators (or states) in short supermultiplets whose scaling dimension does not vary on the conformal manifold. We will use indices $K,L,\ldots$ to label the operators/states in this sector, or equivalently the different directions on the fiber of this vector bundle. As in the previous subsection \ref{radial} we work on a local patch of $\MM$ parametrized by exactly marginal couplings $\lambda^\mu$.

We further assume that the bundles in question are endowed with a connection $\nabla$, a metric $g$ and an operation $C:\VV \otimes \VV \to \VV$. In a local coordinate system the components of $g$ are denoted as $g_{IJ}$ and the components of $C$ as $C_{KL}^M$. In this paper, $\nabla$ is the Berry connection on the states that make up $\VV$, $g_{IJ}=\langle I | J \rangle$ is the inner product of these states, and $C_{KL}^M$ are OPE coefficients in the CFT. 

By construction $\nabla$ is assumed to be compatible with the metric $g$, i.e.\ $\nabla_\mu g_{IJ}=0$. In addition, we assume
\beq
\label{daaa}
\nabla_\mu C_{KL}^M =0
~,
\eeq
which is a non-renormalization theorem of 3-point functions in the protected subsector of interest. 

From equations \eqref{daaa} alone we obtain the following integrability conditions 
\beq
\label{daaba}
[\nabla_\mu, \nabla_{\nu} ] C_{KL}^M =0 ~~\Leftrightarrow~~
({F}_{\mu \nu})_K^S \, C_{SL}^M + ({F}_{\mu \nu})_L^S \, C_{KS}^M 
- ({F}_{\mu \nu})_S^M \, C_{KL}^S
=0
~.
\eeq
A more special version of this identity for $K=L$ reads (no summation over $K$ is assumed)
\beq
\label{daabaa}
({F}_{\mu \nu})_K^S \, C_{KS}^M 
= \frac{1}{2} ({F}_{\mu \nu})_U^M \, C_{KK}^U
~,
\eeq
where we used $C_{KS}^M=C_{SK}^M$. It follows trivially that an arbitrary number of covariant derivatives of ${F}$ also obeys the equations \eqref{daaba}. 

In our context the states on $\VV$ are naturally graded by an R-symmetry charge and the OPE coefficients $C$ always relate states of different R-charges. Consequently, the components of the curvature that appear in  \eqref{daaba}, or \eqref{daabaa}, are always in subspaces of $\VV$ with at least two different values of R-charge.

Without further input the integrability equations \eqref{daaba} do not have a unique solution. The simplest solution is the trivial one
\beq
\label{daabab}
\left({F}_{\mu\nu} \right)_K^L=0
~,
\eeq
but not every theory satisfies \eqref{daaba} in this manner. As we reviewed in the introduction \eqref{daabab} is what happens for 1/2-BPS Coulomb-branch chiral primaries in $\NN=4$ SYM theory.

Another potential solution, which like \eqref{daabab} does not involve further constraints on the OPE coefficients, is
\beq
\label{daaj}
({F}_{\mu \nu})_K^L = \RR_K \,  \delta_K^L \, g_{\mu\nu}
\eeq
with the proviso that the constant coefficients $\RR_K$ are such that
\beq
\label{daak}
\RR_K + \RR_L = \RR_M
\eeq
when $C_{KL}^M \neq 0$. This requirement could be satisfied, for example, if $\RR_K$ are proportional (with a fixed proportionality constant) to an R-symmetry charge that grades the operators/states of interest additively. Notice that \eqref{daaj}, being proportional to the Zamolodchikov metric on $\MM$, obeys automatically $\nabla\, {F} = 0$, i.e.\ the curvature is covariantly constant. As a result, the curvature satisfies trivially the higher-derivative versions of \eqref{daaba}. We are not aware of a CFT that realizes \eqref{daaj}.

Yet another more involved type of solution occurs in the case of CFTs with homogeneous operator bundles, namely bundles of covariantly constant curvature on homogeneous conformal manifolds. It is known that the connection on homogeneous bundles is fully determined by the connection of the base space \cite{Helgason}. An example of this type is exhibited by 1/2-BPS chiral primary bundles in 2d $\NN=(4,4)$ superconformal manifolds. A detailed discussion of this case with explicit formulae for the curvature can be found in \cite{deBoer:2008ss}. In that example the conformal manifold is a coset space of the form 
\beq
\label{daala}
\frac{SO(4,n)}{SO(4) \times SO(n)}
~.
\eeq
In a vielbein basis where the coordinates of the tangent space are labeled by $\mu=({\bf a},{\bf I})$, with $\bf a$ and $\bf I$ indices in the vector representation of $SO(4)$ and $SO(n)$ respectively, the curvature of the chiral primary bundles in a representation ${\mathfrak R}$ of $SO(n)$ takes the form
\beq
\label{daalb}
({F}_{\mu \nu})_K^L = - f \delta_{\bf a \bf b} \left( \Sigma^{\mathfrak R}_{\bf I \bf J} \right)_K^L 
\eeq
for $\mu=(\bf a, \bf I)$, $\nu = (\bf b, \bf J)$. $f$ is a numerical constant and $\Sigma^{\mathfrak R}_{\bf I \bf J}$ a matrix in the representation $\mathfrak R$. Inserting \eqref{daalb} into the integrability equation \eqref{daabaa} we obtain 
\beq
\label{daalc}
\left( \Sigma^{\mathfrak R}_{\bf I \bf J} \right)_K^S C_{KS}^M = \frac{1}{2} \left( \Sigma^{{\mathfrak R}'}_{\bf I \bf J} \right)_U^M C^U_{KK}
~.
\eeq  
The LHS involves the representation $\mathfrak R$ that contains the state labeled by $K$ and the RHS involves the representation ${\mathfrak R}'$ that contains the state labeled by $M$, which is in the tensor product of $\mathfrak R \times \mathfrak R$. It would be interesting to study the relations \eqref{daalc} in more detail.

\section{Curvature of Higgs-branch superconformal primary bundles}
\label{n2}

In this section we focus on four-dimensional $\NN=2$ SCFTs. A large class of such theories with non-trivial superconformal manifolds is provided by the class-$S$ construction in \cite{Gaiotto:2009we}. In what follows we compute the Berry phase of Higgs-branch superconformal primary bundles in general $\NN=2$ SCFTs.

\subsection{Higgs-branch superconformal primaries}
\label{shortening}

The $\NN=2$ superconformal algebra possesses eight Poincar\'e supercharges $Q^{\mathfrak i}_\alpha$, $\bar Q_{{\mathfrak i}\dot \alpha}$ and their superconformal partners $S_{\mathfrak i}^\alpha$, $\bar S^{{\mathfrak i}\dot\alpha}$. The indices ${\mathfrak i}=1,2$ are $SU(2)_R$ indices and the indices $\alpha, \dot\alpha=\pm$ are spacetime spinor indices. We will follow closely the conventions in Ref.\ \cite{Dolan:2002zh}, which are also summarized in appendix D.3 of \cite{Baggio:2017aww}. For the convenience of the reader we have also collected the main definitions in appendix \ref{defs}. In these conventions, after a suitable similarity transformation the odd superconformal generators are denoted as
\beq
\label{n2aa}
\QQ_\alpha^{+{\mathfrak i}} ~, ~~
\QQ^{-\alpha}_{{\mathfrak i}} ~, ~~
\SS^{-{\mathfrak i}\dot\alpha} ~, ~~
\SS^+_{{\mathfrak i}\dot\alpha} 
~.
\eeq
The kets $|I\rangle$ of $SU(2)_R$ highest weight Higgs-branch superconformal primary states obey the following shortening conditions
\beq
\label{n2ab}
\QQ^{+1}_\alpha |I\rangle~, ~~ \QQ^{-\alpha}_{\mathfrak i} |I\rangle =0~, ~~
\SS^+_{2\dot \alpha} |I\rangle =0~, ~~ \SS^{-{\mathfrak i} \dot\alpha} | I\rangle =0~, ~~ ({\mathfrak i}=1,2)
\eeq
from which one can also deduce
\beq
\label{n2ac}
(H-2R)|I \rangle =0
~.
\eeq
$H$ is the conformal Hamiltonian and $R$ the eigenvalue of the $R_3$ Cartan generator of the $SU(2)_R$. For the bras $\langle \bar I |$ the corresponding conditions are
\beq
\label{n2ad}
\langle \bar I | \QQ_1^{-\alpha} =0~, ~~ \langle \bar I | \QQ^{+{\mathfrak i}}_\alpha =0 ~, ~~
\langle \bar I | \SS^{-2\dot\alpha} =0~, ~~ \langle \bar I | \SS^{+}_{{\mathfrak i} \dot \alpha} =0~, ~~ ({\mathfrak i}=1,2)
~.
\eeq
In this notation the 2-point functions are $g_{I\bar J} = \langle \bar J | I \rangle$. We introduced a bar in the notation of the bras to remind us that the corresponding operators are lowest weight vectors in the $SU(2)_R$ representation.

\subsection{Berry curvature of R-symmetry highest weight states}

Our goal here is to evaluate the Berry curvature 
\beq
\label{n2ba}
\left( {F}_{\mu\nu} \right)_{I \bar J} = \sum_{n\notin \HH_I} \sum_{a, \bar b \in \HH_n} \frac{1}{(\Delta_I - \Delta_n )^2} \langle \bar J | \p_\mu H | n,a\rangle g_{(n)}^{\bar b a} \langle n,\bar b | \p_\nu H | I \rangle - (\mu \leftrightarrow \nu)
\eeq 
for the Higgs-branch superconformal primary states \eqref{n2ab}, \eqref{n2ad}. The first task is to find a suitable expression for the Hamiltonian deformations parametrized by the exactly marginal couplings. In \cite{Baggio:2017aww} the $\NN=2$ exactly marginal deformations were expressed in the form
\beq
\label{n2bb}
\delta H = \frac{\delta \lambda^k}{(2\pi)^2} \left( \SS^- \right)^4 \cdot \sint \phi_k + \frac{\delta \bar\lambda^\ell}{(2\pi)^2} \left( \SS^+\right)^4 \cdot \sint \bar\phi_\ell
~,
\eeq
where $\phi_k$ are (Coulomb-branch) $\NN=2$ chiral primaries with $U(1)_R$ charge 4 and the notation $(\SS^\pm)^4 \cdot$ denotes the action of a nested (anti)commutator of the superconformal generators. $\sint$ is shorthand notation for integrals on $S^3$ at fixed (Euclidean) time $\time=0$. This expression worked very efficiently in the computation of the Berry curvature in the $\NN=2$ Coulomb-branch chiral ring \cite{Baggio:2017aww}. Direct computation shows, however, that when applied to the Higgs-branch sector this approach leads to expressions that are much harder to manipulate. Essentially, because of the defining properties of the Higgs-branch superconformal primary states \eqref{n2ab}, \eqref{n2ad} we cannot achieve the complete annihilation of the external states by the supercharges $\SS^\pm$ that appear in \eqref{n2bb}. As a result, the steps that allow us to move $\SS^\pm$ around in the computation of the Berry curvature in the $\NN=2$ Coulomb-branch chiral ring do not work as nicely in the Higgs-branch context. We need a more efficient approach.

In appendix \ref{HamiltonDefs} we explain that there is an equivalent alternative way to write $\delta H$, where two of the $\SS^-$'s on the first term on the RHS of \eqref{n2bb} are replaced by two $\QQ^+$'s and two of the $\SS^+$'s on the second term are replaced by two $\QQ^-$'s. Specifically, we re-express \eqref{n2bb} as
\beq
\label{n2bc}
\delta H = \frac{\delta \lambda^k}{(2\pi)^2} \left( \QQ^{+1} \right)^2 \left( \SS^{-2} \right)^2 \cdot \sint \phi_k + \frac{\delta \bar\lambda^\ell}{(2\pi)^2} \left( \QQ^-_1 \right)^2 \left( \SS^+_2\right)^2 \cdot \sint \bar\phi_\ell
~.
\eeq
As a check of this formula we notice that both interactions on the RHS of \eqref{n2bc} are $U(1)_r$ and $SU(2)_R$ neutral. Moreover, the order of the acting supercharges in each term does not matter because the $\QQ^+$'s commute with the $\SS^-$'s and the $\QQ^-$'s commute with the $\SS^+$'s.

\subsubsection{Summary of the final result}

Inserting \eqref{n2bc} into the general expression \eqref{n2ba}, regularizing and appropriately manipulating with the use of superconformal Ward identities we finally obtain the following result
\begin{subequations}
\beq
\label{n2ca}
\left( {F}_{k\ell} \right)_{I\bar J} = 0~, ~~
\left( {F}_{\bar k\bar \ell} \right)_{I\bar J} = 0
~,
\eeq
\bea
\label{n2cb}
\left( {F}_{k\bar \ell} \right)_{I\bar J} =  
\sum_{\Delta_A=2R_I \atop r_A=-4 \, ,\, R_A=R_I } C_{kA\bar J}g^{\bar B A} C^*_{\bar B \bar \ell I}
- \sum_{\Delta_A=2R_I \atop r_A=4 \, , \, R_A=R_I} C^*_{\bar J \bar \ell A} g^{\bar B A} C_{k I \bar B}
~.
\eea 
\end{subequations}
The details of the computation that leads to these formulae are presented separately in the following subsection \ref{n2details}. In the rest of this subsection we would like to discuss the immediate implications of \eqref{n2ca}, \eqref{n2cb}.

The first two equations in \eqref{n2ca} verify that the Higgs-branch superconformal primary bundles are holomorphic. Equation \eqref{n2cb} expresses the remaining components of the curvature in terms of 2- and 3-point functions in a form that resembles the $[C,\bar C]$ term in the \tt\ equations that express the Coulomb-branch result \cite{Papadodimas:2009eu} (compare with the RHS of eq.\ \eqref{setag} for the corresponding terms in $\NN=4$ SYM theory). There are, however, several important differences between the Higgs-branch and the Coulomb-branch results. 

First, we note that there is no term on the RHS of \eqref{n2cb} proportional to the Zamolodchikov metric (i.e.\ a term like the last term on the RHS of \eqref{setag}). 

Second, the 3-point functions that appear in \eqref{n2cb} involve operators/states from three different superconformal multiplets: $(i)$ the Coulomb-branch chiral primaries with $U(1)_r$ charge $r=\pm 4$ that are responsible for the exactly marginal deformations in \eqref{n2bc}, $(ii)$ the external Higgs-branch superconformal primaries whose Berry curvature we are computing, and $(iii)$ degenerate intermediate states $|A\rangle $, $|B\rangle$ with $U(1)_r$ charges $r=\mp 4$ and the same scaling dimension and $SU(2)_R$ charge as the external Higgs-branch superconformal primaries. A thorough examination of the quantum numbers of states in all possible superconformal multiplets of the $\NN=2$ superconformal algebra \cite{Dolan:2002zh,Cordova:2016emh} reveals that the only states with the above-prescribed selection rules are scalar $(\SS_2^+)^2$ descendants of superconformal primaries in $B_1 \bar L[j=0;\bar j=0]_{\Delta=2R+2}^{(R=R_I-1;r=-4)}$ for the first term on the RHS of \eqref{n2cb} and scalar $(\QQ^{+1})^2$ descendants of superconformal primaries $L\bar B_1[j=0;\bar j=0]_{\Delta=2R+2}^{(R=R_-1;r=4)}$ for the second term.\footnote{In the notation of \cite{Dolan:2002zh} the $B_1 \bar L[j=0; \bar j= 0]_\Delta^{(R;r)}$, $L\bar B_1[j=0;\bar j=0]_\Delta^{(R;r)}$ states are respectively the $\BB_{R,-\frac{r}{2}(0,0)}$, $\bar \BB_{R,-\frac{r}{2}(0,0)}$ states.} The superconformal primary states in $B_1\bar L$ and $L\bar B_1$ are 1/4-BPS mixed Higgs-Coulomb-branch states. The corresponding operators appear naturally in the OPE of Coulomb-branch and Higgs-branch primary states \cite{Argyres:2015ffa}. In parenthesis, we note that operators in the superconformal multiplets $B_1 \bar L$, $L\bar B_1$ can also appear in the OPE of two Coulomb-branch operators \cite{Beem:2014zpa}.

The contributions to the first term on the RHS of eq.\ \eqref{n2cb} involve the 3-point functions 
\beq
\label{n2cc}
C_{kA\bar J} = \langle \bar J | (\SS_2^+)^2 \cdot \AA | k\rangle 
~,
\eeq
which clearly vanish since $\SS_{2\dot\alpha}^+ |k\rangle =0$, $\langle \bar J | \SS_{2\dot\alpha}^+ =0$. $|k\rangle$ is the state created by the Coulomb-branch chiral primary operator $\phi_k$. Similarly, the second term on the RHS of \eqref{n2cb} involves the 3-point functions
\beq
\label{n2cca}
C^*_{\bar J \bar \ell A} = \langle \bar J | (\QQ^{+1})^2 \cdot \AA | \bar \ell \rangle
~,
\eeq
which vanish because $\QQ^{+1}_\alpha |\bar \ell \rangle =0$, $\langle \bar J | \QQ^{+1}_\alpha =0$. $|\bar\ell\rangle$ is the state created by the Coulomb-branch antichiral primary operator $\bar\phi_{\ell}$. We can argue the same results on $\IR^4$ using corresponding superconformal Ward identities.

The conclusion is that both terms on the RHS of eq.\ \eqref{n2cb} vanish automatically, and therefore $\left( {F}_{k\bar \ell} \right)_{I\bar J}=0$ identically in this case. This proves that all the components of the Berry curvature vanish identically on an $\NN=2$ superconformal manifold in the case of the Higgs-branch operators,
\beq
\label{n2cd}
\left( {F}_{\mu\nu } \right)_{I\bar J} =0
~.
\eeq
Hence, for Higgs-branch superconformal primaries there is (at least locally on the superconformal manifold) a coupling-constant-independent basis whether the non-renormalization theorem of 3-point functions takes the form $\p_\mu C_{KLM}=0$.

Unlike the case of the 1/2-BPS chiral primary operators in $\NN=4$ SYM theory, there is no immediate evidence in the above result that the vanishing curvature of the Higgs-branch superconformal primaries is related to a structure of non-trivial identities between correlation functions.

We anticipate similar arguments to apply to the more general Schur operators, but we will not examine their case in detail in this paper.

\subsubsection{Details of the computation}
\label{n2details}

The details of the computation that leads to the central result \eqref{n2ca}-\eqref{n2cb} are summarized here.

Starting from the case where both indices $\mu,\nu$ are holomorphic $(\mu=k,\nu=\ell)$ we first notice that it is convenient to re-express the corresponding curvature components as a limit on an auxiliary parameter $x$ \footnote{A similar manipulation was extremely useful also in the computation of the Berry curvature of Coulomb branch chiral primary states in \cite{Baggio:2017aww}. Several details of the manipulation here are different from the ones in \cite{Baggio:2017aww}.}
\beq
\label{n2df}
\left( {F}_{k\ell}\right)_{I\bar J} = \lim_{x\to 0} \left( \widetilde{F}_{k\ell} \right)_{I\bar J} 
~,
\eeq
where
\bal
\label{n2dg}
\left( \widetilde{F}_{k\ell} \right)_{I\bar J} &:= \frac{1}{(2\pi)^4}
\langle \bar J | \left( \QQ^{+1} \right)^2 \left( \SS^{-2} \right)^2 \cdot \sint \phi_k \left( H-2\RR^1_{~1}-x\right)^{-1} \left( H+2\RR^2_{~2}-x\right)^{-1}
\\
&~~~~~~~~~~~~~~~~~~~\left( \QQ^{+1} \right)^2 \left( \SS^{-2} \right)^2 \cdot \sint \phi_\ell |I \rangle - (k\leftrightarrow \ell)
~.
\end{split}
\end{align}
$\RR^{\mathfrak i}_{~\mathfrak j}$ denotes different $SU(2)_R\times U(1)_r$ generators (see the appendix \ref{defs} for the relevant conventions as well as \cite{Dolan:2002zh}\footnote{$\hat R$ in \cite{Dolan:2002zh} is $-r/2$ in our notation.}). In particular, 
\beq
\label{n2dga}
\RR^1_{~1}= R - \frac{r}{4}~, ~~ \RR^2_{~2} = -R -\frac{r}{4}
~.
\eeq
To keep the notation short from now on we set
\beq
\label{n2dgb}
\HH_1 := H - 2 \RR^1_{~1} = H -2R + \frac{r}{2}~,~~
\HH_2 := H +2 \RR^2_{~2} = H -2R - \frac{r}{2}
~.
\eeq
 
For the next step we use the shortening conditions on the external states and the superconformal algebra relations 
\beq
\label{n2dia}
[ \HH_1, \SS^{-2} ] =0~, ~~ 
[ \HH_2, \SS^{-2} ] = \SS^{-2}
\eeq
to obtain
\bal
\label{n2dj}
&\langle \bar J | \{ \SS^{-2},\Lambda \} (\HH_1-x)^{-1} (\HH_2-x)^{-1}  \left( \QQ^{+1} \right)^2 \left( \SS^{-2} \right)^2 \cdot \sint \phi_\ell | I \rangle 
\\
&=  \langle \bar J | \Lambda \, \SS^{-2} (\HH_1-x)^{-1} (\HH_2-x)^{-1}  \left( \QQ^{+1} \right)^2 \left( \SS^{-2} \right)^2 \cdot \sint \phi_\ell | I \rangle
\\
&=   \langle \bar J | \Lambda (\HH_1-x)^{-1} (\HH_2-1-x)^{-1}  \SS^{-2}  \left[  \left( \QQ^{+1} \right)^2 \left( \SS^{-2} \right)^2 \cdot \sint \phi_\ell \right] | I \rangle 
=0
~.
\end{split}
\end{align}
In these expressions we suppressed the spinor indices and defined the fermionic operator $\Lambda$ so that $\{ \SS^{-2},\Lambda\}=\left( \QQ^{+1} \right)^2 \left( \SS^{-2} \right)^2 \cdot \sint \phi_k$. In this manner we conclude that $(\widetilde {F}_{k\ell} )_{I\bar J}=0$. A similar result holds for the complex conjugate version with $\mu=\bar k, \nu = \bar \ell$. As a result, we derive the holomorphicity conditions \eqref{n2ca} 
\beq
\label{n2dk}
\left( {F}_{k\ell}\right)_{I\bar J} = 0~, ~~
\left( {F}_{\bar k \bar \ell}\right)_{I\bar J} = 0
~.
\eeq

Next, we compute the holomorphic-antiholomorphic components
\beq
\label{n2dl}
\left( {F}_{k\bar \ell} \right)_{I\bar J} = \lim_{x\to 0} \left( \widetilde {F}_{k\bar \ell}\right)_{I\bar J}
~,\eeq
where
\bal
\label{n2dm}
\left( \widetilde{F}_{k\bar \ell} \right)_{I\bar J} &:=
\frac{1}{(2\pi)^4} \langle \bar J |  \left( \QQ^{+1} \right)^2 \left( \SS^{-2} \right)^2 \cdot \sint \phi_k  (\HH_1-x)^{-1} (\HH_2-x)^{-1} \left( \QQ^-_1 \right)^2 \left( \SS^+_2\right)^2 \cdot \sint \bar\phi_\ell | I \rangle 
\\
&- \frac{1}{(2\pi)^4} \langle \bar J | \left( \QQ^-_1 \right)^2 \left( \SS^+_2\right)^2 \cdot \sint \bar \phi_\ell  (\HH_1-x)^{-1} (\HH_2-x)^{-1}  \left( \QQ^{+1} \right)^2 \left( \SS^{-2} \right)^2 \cdot \sint \phi_k | I \rangle 
~.
\end{split}
\end{align}
Consider each line in \eqref{n2dm}. We notice that we can move $\left( \QQ^{+1} \right)^2$, $\left( \SS^{-2} \right)^2$, $\left( \QQ^-_1 \right)^2$, $\left( \SS^+_2\right)^2$ across $H-2R$ with appropriate shifts. For the first line, since $\SS^{+}_{2\dot\alpha}|I\rangle =0$, $\QQ^{-\alpha}_{1}|I\rangle =0$, $\langle \bar J | \SS^{-2\dot\alpha}=0$, $\langle \bar J | \QQ^{+1}_{\alpha}=0$ we can move $(\SS^+_2)^2$ to the left and $(\QQ^{+1})^2$ to the right to obtain
\bal
\label{n2dn}
&\langle \bar J | \left( \QQ^{+1} \right)^2 \left( \SS^{-2} \right)^2 \cdot \sint \phi_k   (\HH_1-x)^{-1} (\HH_2-x)^{-1} \left( \QQ^-_1 \right)^2 \left( \SS^+_2\right)^2 \cdot \sint \bar\phi_\ell | I \rangle
\\
&= \langle \bar J | \sint \phi_k \left[ \left( \QQ^{+1} \right)^2 \left( \SS^{-2} \right)^2  (\HH_1-x)^{-1} (\HH_2-x)^{-1} \left( \QQ^-_1 \right)^2 \left( \SS^+_2\right)^2 \right] \sint \bar\phi_\ell | I \rangle
\\
&= \langle \bar J | \sint \phi_k \left[ \left( \SS^{-2} \right)^2 \left( \SS^{+}_2 \right)^2    (\HH_1+2-x)^{-1} (\HH_2-2-x)^{-1}  \left( \QQ^{+1} \right)^2 \left( \QQ^-_1\right)^2 \right] \sint \bar\phi_\ell | I \rangle
\\
&= \langle \bar J | \sint \phi_k \left[ (\SS^{-2})^2, ( \SS^+_2)^2 \right]  (\HH_1+2-x)^{-1} (\HH_2-2-x)^{-1} \left[ (\QQ^{+1})^2, ( \QQ^-_1)^2 \right] \sint \bar\phi_\ell | I \rangle
.
\\
\end{split}
\end{align}
For the second line a similar computation gives
\begin{align}
\begin{split}
\label{n2do}
&\langle \bar J | \left( \QQ^-_1 \right)^2 \left( \SS^+_2\right)^2 \cdot \sint \bar\phi_\ell  (\HH_1-x)^{-1} (\HH_2-x)^{-1} \left( \QQ^{+1} \right)^2 \left( \SS^{-2} \right)^2 \cdot \sint \phi_k | I \rangle
\\
&= \langle \bar J | \sint \bar\phi_\ell \left[ (\QQ^{+1})^2,(\QQ^-_1 )^2 \right]  (\HH_1-2-x)^{-1} (\HH_2+2-x)^{-1} \left[ (\SS^{-2})^2,(\SS^+_2 )^2 \right] \sint \phi_k | I \rangle
.
\end{split}
\end{align}
Consequently, we arrive at the expression
\bal
\label{n2dp}
\left( \widetilde{F}_{k\bar \ell} \right)_{I\bar J} =&
\frac{1}{(2\pi)^4} \langle \bar J | \sint \phi_k \frac{  \left[ (\SS^{-2})^2,(\SS^+_2 )^2 \right] \left[ (\QQ^{+1})^2,(\QQ^-_1 )^2 \right] }{(\HH_1+2-x) (\HH_2-2-x) } \sint \bar\phi_\ell | I \rangle
\\
&- \frac{1}{(2\pi)^4}\langle \bar J | \sint \bar\phi_\ell \frac{ \left[ (\QQ^{+1})^2,(\QQ^-_1 )^2 \right] \left[ (\SS^{-2})^2,(\SS^+_2 )^2 \right] }{(\HH_1-2-x) (\HH_2+2-x) }  \sint \phi_k | I \rangle
~.
\end{split}
\end{align}

A straightforward application of the (anti)commutation relations of the superconformal algebra gives\footnote{The generators $\tilde M_{\alpha}^{~\beta}$ and $\tilde{\bar M}^{\dot \beta}_{\dot\alpha}$ are directly related to the Lorentz generators. We refer the reader to \cite{Dolan:2002zh} for the appropriate conventions (see also appendix \ref{defs}). Moreover, we use the conventions $\varepsilon_{12}=-\varepsilon^{12}=-1$.}
\begin{subequations}
\begin{align}
\begin{split}
\label{n2dqa}
\left[ (\QQ^{+1})^2,(\QQ^-_1 )^2 \right] &= - \frac{4}{16} \varepsilon^{\alpha\beta} \varepsilon_{\gamma \delta} 
\bigg[ \left( \delta_\alpha^\gamma \HH_1 
+ 2 \tilde M_\alpha^{~\gamma}\right)
\left( \QQ_\beta^{+1} \QQ_1^{-\delta} - \QQ_1^{-\delta} \QQ_\beta^{+1} \right)
\\
&~~~~~~~~~~~~~~
+ 4 \delta_\beta^\delta  \left( \delta_\alpha^\gamma \HH_1
+ 2 \tilde M_\alpha^{~\gamma}\right) \bigg]
 ~,
\end{split}
\end{align}
or
\begin{align}
\begin{split}
\label{n2dqab}
\left[ (\QQ^{+1})^2,(\QQ^-_1 )^2 \right] &= + \frac{4}{16} \varepsilon^{\alpha\beta} \varepsilon_{\gamma \delta} 
\bigg[ 
\left(- \QQ_\beta^{+1} \QQ_1^{-\delta} + \QQ_1^{-\delta} \QQ_\beta^{+1} \right)
\left( \delta_\alpha^\gamma \HH_1 
+ 2 \tilde M_\alpha^{~\gamma}\right)
\\
&~~~~~~~~~~~~~~ - 4 \delta_\beta^\delta  \left( \delta_\alpha^\gamma \HH_1  
+ 2 \tilde M_\alpha^{~\gamma}\right) \bigg]
 ~,
\end{split}
\end{align}
and 
\begin{align}
\begin{split}
\label{n2dqb}
\left[ (\SS^{-2})^2,(\SS^+_2 )^2 \right] &= -\frac{4}{16} \varepsilon_{\dot\alpha \dot\beta} \varepsilon^{\dot\gamma \dot \delta} 
\bigg[ \left( \SS^{-2\dot \alpha} \SS^+_{2\dot\gamma} - \SS^+_{2\dot\gamma} \SS^{-2\dot\alpha} \right) \left( \delta^{\dot\beta}_{\dot \delta} \HH_2  
- 2 {\tilde {\bar M}}^{\dot \beta}_{\dot \delta} \right) 
\\
&~~~~~~~~~~~~~~ -4 \delta^{\dot \beta}_{\dot \delta} \left( \delta^{\dot \alpha}_{\dot \gamma} \HH_2 
-2 \tilde{\bar M}^{\dot \beta}_{\dot\delta} \right) \bigg]
~,
\end{split}
\end{align}
or
\begin{align}
\begin{split}
\label{n2dqbb}
\left[ (\SS^{-2})^2,(\SS^+_2 )^2 \right] &= +\frac{4}{16} \varepsilon_{\dot\alpha \dot\beta} \varepsilon^{\dot\gamma \dot \delta} 
\bigg[ 
\left( \delta^{\dot\beta}_{\dot \delta} \HH_2  
- 2 {\tilde {\bar M}}^{\dot \beta}_{\dot \delta} \right) 
\left(- \SS^{-2\dot \alpha} \SS^+_{2\dot\gamma} + \SS^+_{2\dot\gamma} \SS^{-2\dot\alpha} \right) 
\\
&~~~~~~~~~~~~~~ +4 \delta^{\dot \beta}_{\dot \delta} \left( \delta^{\dot \alpha}_{\dot \gamma} \HH_2 
-2 \tilde{\bar M}^{\dot \beta}_{\dot\delta} \right) \bigg]
~.
\end{split}
\end{align}
\end{subequations}
The overall $\frac{1}{16}$ factors are part of the conventions. The terms that involve $\tilde M_{\alpha}^{~\beta}$ do not contribute in \eqref{n2dp}. The same holds for terms with $\QQ_\beta^{+1}$ on the very right and terms with $\SS_{2\dot\alpha}^+$ on the very left in the first line of \eqref{n2dp}. In the second line of \eqref{n2dp} a similar statement applies to terms with $\QQ_\beta^{+1}$ on the very left and terms with $\SS_{2\dot\alpha}^+$ to the very right. Hence, for our purposes $\left[ (\QQ^{+1})^2,(\QQ^-_1 )^2 \right] \simeq \HH_1(\HH_1+2)$ 
when it is on the right and $\left[ (\QQ^{+1})^2,(\QQ^-_1 )^2 \right] \simeq -\HH_1 (\HH_1-2)$ when it is on the left. Similarly, $\left[ (\SS^{-2})^2,(\SS^+_2 )^2 \right] \simeq \HH_2(\HH_2-2)$ when it is on the left and $\left[ (\SS^{-2})^2,(\SS^+_2 )^2 \right] \simeq -\HH_2(\HH_2+2)$ when it is on the right. 

With this information eq.\ \eqref{n2dp} becomes
\bal
\label{n2dt}
\left( \widetilde{F}_{k\bar \ell} \right)_{I\bar J} =&
 \frac{1}{(2\pi)^4} \langle \bar J | \sint \phi_k \frac{\HH_2(\HH_2-2)\HH_1(\HH_1+2)}{(\HH_1+2-x) (\HH_2-2-x)} \sint \bar\phi_\ell | I \rangle
\\
&- \frac{1}{(2\pi)^4}\langle \bar J | \sint \bar\phi_\ell \frac{\HH_1(\HH_1 {-}2)\HH_2(\HH_2{+}2)}{(\HH_1-2-x) (\HH_2+2-x) }  \sint \phi_k | I \rangle
~.
\end{split}
\end{align}
Using the shortening conditions $\HH_1 | I \rangle = 0$, $\HH_2 | I \rangle =0$ and the commutation relations 
\beq
\label{n2dub}
[ \HH_1,\phi ] = (\p_\time +2) \phi~, ~~
[ \HH_1,\bar\phi] = (\p_\time -2) \bar\phi~, 
\eeq
\beq
\label{n2duc}
[ \HH_2,\phi ] = (\p_\time -2) \phi~, ~~
[ \HH_2,\bar\phi] = (\p_\time +2) \bar\phi
\eeq
for $\Delta=2$ Coulomb branch (anti)chiral primaries $\phi$ ($\bar\phi$), we find
\bal
\label{n2duda}
\left( \widetilde{F}_{k\bar \ell} \right)_{I\bar J} =  
\frac{1}{(2\pi)^4} \bigg[&
\langle \bar J | \sint \phi_k \frac{(\p_\time+2) \p_\time (\p_\time-2) \p_\time}{(\p_\time -x)^2} \sint \bar\phi_\ell   |I \rangle 
\\
& - \langle \bar J | \sint \bar\phi_\ell \frac{(\p_\time +2)\p_\time (\p_\time-2)\p_\time}{(\p_\time -x)^2} \sint \phi_k | I \rangle \bigg]
~.
\end{split}
\end{align}
After taking the $x\to 0$ limit
\bea
\label{n2dud}
\left( {F}_{k\bar \ell} \right)_{I\bar J} = {\rm (subtracted)} 
+ \frac{1}{(2\pi)^4} \bigg[
\langle \bar J | \sint \varphi_k (\p_\time^2 -4)  \sint \bar\varphi_\ell   |I \rangle 
- \langle \bar J | \sint \bar\varphi_\ell (\p_\time^2 -4) \sint \varphi_k | I \rangle \bigg]
~.\nonumber\\
\eea
The term denoted as `(subtracted)' is due to contributions in \eqref{n2dt} that behave as $0/0$ in the limit $x\to 0$ and have to be subtracted. A short inspection of \eqref{n2dt} reveals that
\beq
\label{n2dva}
{\rm(subtracted)} = \sum_{{E_A=E_B=2R_I \atop r_A=r_B=-4, R_A=R_B=R_I}} C_{kA\bar J}g^{\bar B A} C^*_{\bar B \bar \ell I}
-\sum_{{E_A=E_B=2R_I \atop r_A=r_B=4, R_A=R_B=R_I}} C^*_{\bar J \bar \ell A} g^{\bar B A} C_{k I \bar B} 
~.
\eeq
That leaves the remaining, contact terms on the RHS of \eqref{n2dud} that can be evaluated as follows.

The contact term in \eqref{n2dud} involves the insertions of two operators integrated on $S^3$ at the same time $\time=0$. When the operators collide they exhibit UV divergences. These can be regularized with a small relative displacement in time that we have to send to zero at the end of the computation. Following \cite{Baggio:2017aww} we set   
\bal
\label{n2dvb}
\left( \RR_{k\bar l} \right)_{I\bar J} \equiv \frac{1}{(2\pi)^4}
\lim_{\time_1 = -\varepsilon \to 0^-} \bigg[&
\langle \bar J | \sint \phi_k(\time_1) (\p_{\time_2}^2 -4)  \sint \bar\phi_\ell(\time_2)   |I \rangle 
\\
&- \langle \bar J | \sint \bar\phi_\ell(\time_1) (\p_{\time_2}^2 -4) \sint \phi_k(\time_2) | I \rangle \bigg]_{\time_2=0}
~,
\end{split}
\end{align}
which is equal to 
\bal
\label{n2dvba}
\left( \RR_{k\bar l} \right)_{I\bar J} =&
\frac{1}{(2\pi)^4} 
\bigg[ - e^{2(\time_1+\time_2)} \p_{\time_1}\p_{\time_2} 
\left( e^{-2(\time_1+\time_2)} \langle \bar J | \sint \phi_k(\time_1) \sint \bar\phi_\ell(\time_2)   |I \rangle \right)
\\
& + e^{2(\time_1+\time_2)} \p_{\time_1}\p_{\time_2} 
\left( e^{-2(\time_1+\time_2)} \langle \bar J | \sint \bar \phi_l(\time_1) \sint \phi_k(\time_2)   |I \rangle \right)
\bigg]_{\time_1 = -\varepsilon \to 0^-,\time_2=0}
.
\end{split}
\end{align}
It is worth comparing this expression with the properly normalized contact term (7.26) in \cite{Baggio:2017aww}
\bal
\label{n2dvbb}
\left( \widehat \RR_{k\bar l} \right)_{I\bar J} =&
\frac{1}{(2\pi)^4} 
\bigg[ - e^{2(\time_1+\time_2)} \p_{\time_1}\p_{\time_2} 
\left( e^{-4 \time_1} \langle \bar J | \sint \phi_k(\time_1) \sint \bar\phi_\ell(\time_2)   |I \rangle \right)
\\
& + e^{2(\time_1+\time_2)} \p_{\time_1}\p_{\time_2} 
\left( e^{-4 \time_2} \langle \bar J | \sint \bar \phi_l(\time_1) \sint \phi_k(\time_2)   |I \rangle \right)
\bigg]_{\time_1 = -\varepsilon \to 0^-,\time_2=0}
\end{split}
\end{align}
that expresses the contact term in the curvature of the Coulomb-branch chiral primaries.

Transforming back to the plane with the change of coordinates $r\equiv |x| = e^{\time_1}$, $\rho \equiv |y| = e^{\time_2}$, $\varphi(\time_1) = r^2 \varphi(x)$, etc., and using the conformal block expansion
\beq
\label{n2dbbc}
\langle \bar J | \phi_k (x) \bar \phi_\ell (y) | I\rangle = \frac{1}{|x-y|^4} \sum_\OO C_{k\bar l}^\OO C_{\OO I\bar J} g_{\Delta,\ell}(u,v)
\eeq
we obtain
\beq
\label{n2dbbd}
\left( \RR_{k\bar l} \right)_{I\bar J} = - \sum_\OO C^\OO_{k\bar l} C_{\OO I \bar J} X_{\Delta,\ell}
\eeq
where
\beq
\label{n2dbbe}
X_{\Delta,\ell} = \lim_{r\to 1^-} \int_{|x|=r} d\Omega_3^x \int_{|y|=\rho=1} d\Omega_3^y\, |x|^2 |y|^2 (x\cdot \p_x) (y\cdot \p_y) \left [ \frac{1-(-1)^\ell}{|x-y|^4} g_{\Delta,\ell}(u,v) \right]
~.
\eeq
Immediately we notice that
\beq
\label{n2dbbf}
X_{\Delta, \ell\, =\, even}=0
~.
\eeq
For odd $\ell$ we have
\beq
\label{n2dbbg}
X_{\Delta,\ell} = 2 \lim_{r\to 1^-} \int_{|x|=r} d\Omega_3^x \int_{|y|=\rho=1} d\Omega_3^y\, |x|^2 |y|^2 (x\cdot \p_x) (y\cdot \p_y) \left [ \frac{1}{|x-y|^4} g_{\Delta,\ell}(u,v) \right]
~.
\eeq
Using the known properties of the conformal partial waves $g_{\Delta,\ell}(u,v)$, \cite{Dolan:2000ut}, we can check that there are no values of $\Delta$, $\ell$ for which we get a contribution to \eqref{n2dbbd}. For odd $\ell\geq 3$ $X_{\Delta,\ell}$ vanishes in the limit $r\to 1$. A potential contribution could arise from $\Delta=3$, $\ell=1$, namely the conformal block of the R-symmetry currents $\JJ$. In the case at hand, however, the external Higgs-branch superconformal primary states have vanishing $U(1)_R$ charge and the Coulomb-branch chiral primaries $\phi_k$ have vanishing $SU(2)_R$ charge, hence even this case cannot contribute. 

In summary, we deduce that the contact term $\left( \RR_{k\bar l} \right)_{I\bar J}$ is identically zero. Combining this result with \eqref{n2dva} we recover the advertised formula \eqref{n2cb}.

\subsection{R-symmetry descendants}
\label{n2descendants}

As a minor check of the formalism, in this subsection we compute the Berry curvature for Higgs-branch superconformal primary states in irreducible representations of $SU(2)_R$ that are descendants of a highest weight state. Assuming that the Berry phase of the R-symmetry generators vanishes, we should find that all descendants have the same Berry curvature as the highest weight states. 

Let $|I\rangle$ be a highest weight state in a spin $R$ representation of $SU(2)_R$. $R$ is a positive half-integer. The $SU(2)_R$ generators are $R_\pm, R_3$. We denote the descendants as
\beq
\label{dbbaa}
| I,{\mathfrak m} \rangle := \NN_{\mathfrak m} \left( R_- \right)^{\mathfrak m} |I\rangle~, ~~ {\mathfrak m} = 1,2,\ldots, 2R
~,\eeq
where $\NN_{\mathfrak m}$ is a real normalization constant that we fix in a moment. 
$\mathfrak m=0$ is the case of a highest weight state that was considered previously. It is excluded in this subsection. For the bras we set
\beq
\label{dbbab}
\langle \bar I, {\mathfrak m} | := \NN_{\mathfrak m} \langle \bar I | \left( R_+ \right)^{\mathfrak m}
\eeq
and fix the normalization factor by setting
\beq
\label{dbbaaa}
\NN_{\mathfrak m} = \sqrt{\frac{\langle \bar I | I \rangle }{\langle \bar I | \left( R_+ \right)^{\mathfrak m} \left( R_- \right)^{\mathfrak m} | I \rangle}}
~,\eeq
which guarantees that $\langle \bar I,{\mathfrak m} | I, {\mathfrak m} \rangle = \langle \bar I | I\rangle$.
As an example, we note that a straightforward use of the $SU(2)$ commutation relations gives
$\NN_1 = \frac{1}{\sqrt {2 R}}$.

We argued in section \ref{gendescend} that only states with the same $\mathfrak m$ can mix. Consequently, we can focus on the following quantities
\beq
\label{dbbac}
\left( {F}^{(\mathfrak m)}_{\mu\nu} \right)_{I\bar J} = \sum_{n\notin \HH_I} \sum_{a,\bar b \in \HH_n} \frac{1}{(\Delta_I - \Delta_n )^2} \langle \bar J,{\mathfrak m} | \p_\mu H | n,a\rangle g_{(n)}^{\bar b a} \langle n,\bar b | \p_\nu H | I,{\mathfrak m} \rangle - (\mu \leftrightarrow \nu)
~.
\eeq
Repeating previous steps we first compute the auxiliary quantity
\beq
\label{dbbaea}
\left( \widetilde {F}^{({\mathfrak m})}_{\mu\nu} \right)_{I\bar J}  
= \langle \bar J, {\mathfrak m} | \p_\mu H ( \HH_1-2{\mathfrak m} -x)^{-1} (\HH_2 -2{\mathfrak m} -x )^{-1} \p_\nu H | I,{\mathfrak m} \rangle - (\mu \leftrightarrow \nu)
\eeq
whose $x\to 0$ limit recovers $\left( {F}_{\mu\nu} \right)_{I\bar J}$ in \eqref{dbbac}. Using the commutation relations 
\beq
\label{dbbaeb}
\left[ R_+, \left( \QQ^{+1} \right)^2 \left( \SS^{-2} \right)^2 \cdot \phi_k \right] =0~, ~~
\left[ R_+, \left( \QQ^{-}_1 \right)^2 \left( \SS^+_2 \right)^2 \cdot \bar \phi_\ell \right] =0
\eeq
and
\beq
\label{dbbaec}
\left[ R_+, \HH_1 \right] = 2 R_+~, ~~
\left[ R_+, \HH_2 \right] = 2 R_+
\eeq
we can easily show that
\begin{align}
\begin{split}
\label{dbbaee}
\left( \widetilde {F}^{({\mathfrak m})}_{\mu\nu} \right)_{I\bar J}  
&= \NN_{\mathfrak m}^2 
\langle \bar J | \p_\mu H ( \HH_1 -x)^{-1} (\HH_2 -x )^{-1} \p_\nu H \left( R_+\right)^{\mathfrak m} \left( R_- \right)^{\mathfrak m} | I \rangle - (\mu \leftrightarrow \nu)
\\
&= \langle \bar J | \p_\mu H ( \HH_1 -x)^{-1} (\HH_2 -x )^{-1} \p_\nu H | I \rangle - (\mu \leftrightarrow \nu)
~,
\end{split}
\end{align}
which coincides with the expression for ${\mathfrak m}=0$.

Alternatively, we could try to compute \eqref{dbbac} from scratch using the logic of the previous subsections. This approach is more involved.

\section{Open problems}
\label{n4}
 
For the reasons outlined in the introduction, it is of interest to explore more generally how the curvature of protected operator bundles behaves. As we emphasized, when non-renormalization theorems exist, they constrain the dynamics of the corresponding sectors, but do not necessarily lead to trivial Berry phases and trivial connections on the superconformal manifold. There can be an interesting interplay between the non-renormalization theorems and the Berry curvature, as we argued for example in section \ref{nonrenormImplications}, but it is not clear, in general, what to expect for the Berry curvature in the presence of a non-renormalization theorem. Given what we know from examples of protected sectors with non-renormalization theorems one might be tempted to conjecture that the curvature of such sectors is always covariantly constant. It would be interesting to prove, or disprove this expectation.

In section \ref{n2} we considered the curvature of Higgs-branch superconformal primary bundles in 4d $\NN=2$ theories. In this example we found that the curvature vanishes identically. We anticipate that the computation of section \ref{n2} extends to the case of arbitrary Schur operators and that the curvature of these operators vanishes as well. It would be useful to spell out the details.

Another, seemingly more involved, example that would be interesting to study is the 1/4-BPS sector in $\NN=4$ SYM theory. For a concise list of the short multiplets in $\NN=4$ SYM theory we refer the reader to \cite{Dolan:2002zh}, or the more recent \cite{Cordova:2016emh}. The superconformal primary operators in these multiplets are labelled by the three Dynkin labels $(h_1,h_2,h_3)$ of the $SU(4)_R$ R-symmetry group. An $SU (4)_R$ highest weight representation with highest weight $(h_1,h_2,h_3)$ is usually denoted as $[h_1,h_2,h_3]$. It is well-known that the scalar conformal primary operators $\phi_I $ in the representation $[q,p,q]$ with scaling dimension $\Delta=p+2q$ are 1/4-BPS for $q\geq 1$ and 1/2-BPS for $q=0$. 

It has been argued that the 2- and 3-point functions of any combination of the above operators do not renormalize. We reviewed the 1/2-BPS operators in the introduction. The 1/4-BPS operators are more subtle \cite{Ryzhov:2001bp,DHoker:2003csh}. The non-renormalization of 2- and 3-point functions of 1/4-BPS operators was shown using analytic superspace arguments in \cite{Heslop:2001gp}. A more elementary argument for the special case of 3-point functions with one 1/4-BPS and two 1/2-BPS insertions was presented in \cite{Baggio:2012rr}.

It would be interesting to determine the curvature of 1/4-BPS superconformal primary bundles and resolve the following questions: Is the curvature of this sector covariantly constant? Can it be expressed solely in terms of data in the 1/4-BPS sector? Is the curvature zero and does this information carry any implications about the structure of correlation functions involving 1/4-BPS operators? Interestingly, one cannot repeat the arguments of appendix \ref{half} to show that the curvature of the 1/4-BPS sector vanishes as we did for the 1/2-BPS sector. That may be an indication that the 1/4-BPS superconformal primary bundles are not flat in which case they would exhibit more intriguing dynamics compared to their 1/2-BPS cousins. We hope to return to these questions in a future publication.

\subsection*{Acknowledgments}

\noindent
I would like to thank Paul Heslop, Kyriakos Papadodimas, Leonardo Rastelli and Balt van Rees for helpful discussions. This work has been supported by STFC under the consolidated grant ST/P000371/1. I would also like to thank the Galileo Galilei Institute for Theoretical Physics for the hospitality and the INFN for partial support during the completion of this work.

\begin{appendix}

\section{Useful definitions}
\label{defs}

In this paper we follow closely the conventions of \cite{Dolan:2002zh}. For the convenience of the reader we collect some of the main definitions in this appendix. 

In the context of radial quantization it is convenient to transform the operators $\OO$ by a similarity transformation 
\beq
\label{defsaa}
\OO \longrightarrow \widetilde \OO = e^{\frac{\pi}{4}(P_0 - K_0)}\, \OO\, e^{-\frac{\pi}{4}(P_0 - K_0)}
~,
\eeq
where $P_0$ and $K_0$ are respectively the time components of the linear momentum and the generator of special conformal transformations. The conformal Hamiltonian $H$ arises from the dilatation operator $D$ as
\beq
\label{defsab}
H = e^{\frac{\pi}{4}(P_0 - K_0)} (- i D ) e^{-\frac{\pi}{4}(P_0 - K_0)}
~.
\eeq

On the plane $\IR^4$ the $\NN=2$ superconformal algebra possesses the Poincar\'e supercharges $Q^{\mathfrak i}_\alpha$, $\bar Q_{{\mathfrak i}\dot\alpha}$ and their superconformal partners $S_{\mathfrak i}^\alpha$, $\bar S^{{\mathfrak i}\dot\alpha}$. The indices ${\mathfrak i}=1,2$ are $SU(2)_R$ indices and the indices $(\alpha,\dot\alpha= \pm)$ are standard spinor indices.

The calligraphic superconformal generators that we use in section \ref{n2} are defined as \cite{Minwalla:1997ka,Dolan:2002zh}
\vspace{-0.4cm}
\begin{subequations}
\label{SCAconventions}
\bea
\label{defsac}
\QQ^{+{\mathfrak i}}_\alpha &=& e^{\frac{\pi}{4}(P_0-K_0)} Q^{\mathfrak i}_\alpha e^{-\frac{\pi}{4}(P_0-K_0)} 
= \frac{1}{\sqrt 2} \left( Q^{\mathfrak i}_\alpha + \sigma_{0\, \alpha\dot\alpha} \bar S^{{\mathfrak i}\dot\alpha} \right)~,
\\
\QQ^{-\alpha}_{\mathfrak i} &=& e^{\frac{\pi}{4}(P_0-K_0)} S_{\mathfrak i}^\alpha e^{-\frac{\pi}{4}(P_0-K_0)}  
= \frac{1}{\sqrt 2} \left( S_{\mathfrak i}^\alpha + \bar Q_{{\mathfrak i}\dot\alpha} \bar \sigma_0^{~\dot\alpha \alpha} \right)~,
\\
- \SS^{-{\mathfrak i}\dot\alpha} &=& e^{\frac{\pi}{4}(P_0-K_0)} \bar S^{{\mathfrak i}\dot\alpha} e^{-\frac{\pi}{4}(P_0-K_0)} 
=\frac{1}{\sqrt 2} \left( \bar S^{{\mathfrak i}\dot\alpha} - \bar\sigma_0^{~\dot\alpha \alpha} Q^{\mathfrak i}_\alpha \right)~,
\\
\label{defsad}
\SS^+_{{\mathfrak i}\dot\alpha} &=& e^{\frac{\pi}{4}(P_0-K_0)} \bar Q_{{\mathfrak i}\dot\alpha} e^{-\frac{\pi}{4}(P_0-K_0)} =\frac{1}{\sqrt 2} \left( \bar Q_{{\mathfrak i}\dot\alpha} - S_{\mathfrak i}^\alpha \sigma_{0 \, \alpha \dot\alpha} \right)
~.
\eea
\end{subequations}
They obey the (anti)commutation relations
\begin{subequations}
\bea
\label{defsae}
\{ \QQ^{+{\mathfrak i}}_{\alpha}, \QQ^{-\beta}_{\mathfrak j} \} &=& 2 \delta^{\mathfrak i}_{~{\mathfrak j}}\delta_{\alpha}^{~\beta} H 
+ 4 \delta^{\mathfrak i}_{~{\mathfrak j}} \tilde M_\alpha^{~\beta}
- 4 \delta_{\alpha}^{~\beta} R^{\mathfrak i}_{~{\mathfrak j}}~,
\\
\{ \SS^{-{\mathfrak i}\dot\alpha}, \SS^+_{{\mathfrak j}\dot\beta} \} &=&  2 \delta^{\mathfrak i}_{~{\mathfrak j}}\delta^{\dot \alpha}_{~\dot \beta} H 
- 4 \delta^{\mathfrak i}_{~{\mathfrak j}} \tilde {\bar M}^{\dot \alpha}_{~\dot \beta}
+ 4 \delta^{\dot \alpha}_{~\dot \beta} R^{\mathfrak i}_{~{\mathfrak j}}~,
\eea
\end{subequations}
and have Hermiticity properties
\beq
\label{defsaf}
\left( \QQ^{+{\mathfrak i}}_\alpha \right)^\dagger = \QQ^{-\beta}_{\mathfrak i} \sigma_{0\beta \dot \alpha}~, ~~
\left( \SS^+_{{\mathfrak i}\dot\alpha} \right)^\dagger = \sigma_{0\alpha\dot\beta} \SS^{-{\mathfrak i}\dot\beta}~.
\eeq
We have adopted the convention $\sigma_0 = \bar \sigma_0= 1$. Furthermore, the rotation generators $\tilde M_{\alpha}^{~\beta}$, $\tilde{\bar M}^{\dot \alpha}_{~\dot\beta}$, which appear in eqs.\ \eqref{n2dqa}-\eqref{n2dqbb}, are defined as the transformation \eqref{defsaa} of the Lorentz generators
\beq
\label{defsag}
M_\alpha^{~\beta} = -\frac{i}{4}\left( \sigma^\mu \bar\sigma^\nu \right)_\alpha^{~\beta} M_{\mu\nu}~, ~~
\bar M^{\dot \alpha}_{~\dot \beta} = -\frac{i}{4}\left( \bar \sigma^\mu \sigma^\nu \right)^{\dot \alpha}_{~\dot \beta} M_{\mu\nu}~.
\eeq
The $\NN=2$ $U(2)_R$-symmetry generators are
\beq
\label{defsai}
\left( R^{\, \mathfrak i}_{~{\mathfrak j}} \right) = \left( {R_3 \atop R_-} ~ {R_+ \atop -R_3} \right) -\frac{r}{4} \left( {1 \atop 0 } ~ {0 \atop 1} \right)~,
\eeq
where $R_\pm, R_3$ are $SU(2)_R$ generators and $r$ is the $U(1)_r$ generator. To simplify the notation, in the main text we frequently use $R$ to denote $R_3$ or its eigenvalue.

For further details we refer the reader to \cite{Dolan:2002zh}.

\section{Hamiltonian deformations}
\label{HamiltonDefs}

In this appendix we collect useful details about the form of the supersymmetric Hamiltonian deformations that are considered in the main text in sections \ref{n2} and \ref{n4}. In particular, we discuss the reasons behind the free exchange of $\QQ^+$ with $\SS^-$ and $\QQ^-$ with $\SS^+$ in the Hamiltonian deformation. This exchange was employed successfully in \cite{Baggio:2017aww} to obtain the \tt\ equations in 4d $\NN=2$ SCFTs and 2d $\NN=(2,2)$ SCFTs. It is further tested in the context of 1/2-BPS operators in 4d $\NN=4$ SYM theory in appendix \ref{half} below.

In a four-dimensional superconformal field theory with $\NN$ supersymmetries\footnote{Here $\NN=2,4$.} there are left-chiral supercurrents $G^{{\mathfrak i}\mu}_\alpha$ and right-chiral superconformal currents $\bar G^{\mathfrak i}_{\dot\alpha}$. Both are conformal primary operators of scaling dimension $\Delta=\frac{7}{2}$. ${\mathfrak i}=1,\ldots,\NN$ are fundamental R-symmetry indices. In what follows we will focus on the left-chiral part of the Poincar\'e supercharges and their right-chiral conformal supercharges. Analogous statements apply obviously to the remaining supercharges.

Conserved currents can be obtained by multiplying with a conformal Killing spinor $\psi^\alpha(x)$
\beq
\label{app4daa}
j^{{\mathfrak i}\mu} = \psi^\alpha(x) G_\alpha^{{\mathfrak i}\mu}(x)
~.
\eeq
The corresponding conserved supercharges are $\int d^3 x \, j^{{\mathfrak i}0}$. On $\IR^4$ the general Killing spinor is
\beq
\label{app4dab}
\psi^\alpha(x) = \lambda^\alpha + x^{\dot \alpha \alpha} \mu_{\dot \alpha}
~,
\eeq
where $\lambda^\alpha$ is an arbitrary constant $(1/2,0)$ spinor, $\mu_{\dot\alpha}$ is an arbitrary constant $(0,1/2)$ spinor and as usual $x^{\dot\alpha \alpha} = x^\mu \sigma_\mu^{\dot\alpha \alpha}$. Hence, in flat space we have the supercharges
\beq
\label{app4dac}
Q^{\mathfrak i}_\alpha \sim \int d^3 x \, G^{\mathfrak i}_{\alpha 0} ~, ~~
\bar S^{{\mathfrak i}\dot\alpha} \sim \int d^3 x\, x^{\dot\alpha \alpha} G_{\alpha 0}^{\mathfrak i}
~.
\eeq

To determine the action of a supercharge on a conformal primary operator $\varphi$ we use the OPE
\beq
\label{app4dad}
G^{{\mathfrak i}\mu}_\alpha (x) \, \varphi(y) = \ldots + \frac{(x-y)^\mu (x-y)_{\alpha \dot \beta} }{2\pi^2 |x-y|^6} \left[ \bar S^{{\mathfrak i}\beta},\varphi \right](y) + \frac{(x-y)^\mu}{2\pi^2 |x-y|^4} \left[ Q^{\mathfrak i}_\alpha, \varphi \right](y) +\ldots
~.\eeq 
We are using notation where $[{\bf Q},\varphi](y)$ means that we first compute the commutator $[{\bf Q},\varphi(0)]$ at the origin and then translate to the point $y$. $[{\bf Q},\varphi(y)]$ denotes the commutator with the operator $\varphi$ at $y$. Clearly, $\left[ \bar S^{{\mathfrak i}\beta},\varphi \right](y) \neq \left[ \bar S^{{\mathfrak i}\beta},\varphi(y) \right]$, but $\left[ Q^{\mathfrak i}_\alpha, \varphi \right](y)= \left[ Q^{\mathfrak i}_\alpha, \varphi (y) \right]$. For the general current \eqref{app4daa}, \eqref{app4dab} a short computation shows that
\beq
\label{app4dae}
\lim \int d^3 x \, j^{{\mathfrak i}0} (x) \, \varphi(y) = \psi^\alpha(y) \left[ Q^{\mathfrak i}_\alpha,\varphi \right] (y) - \p_\mu \psi^\alpha\, \sigma^\mu_{\alpha\dot\beta}(y) \left[ \bar S^{{\mathfrak i}\dot \beta},\varphi \right] (y)
~.
\eeq
The $\lim$ denotes a limit where the $x$-integral is taken close to the $y$ insertion. From here on we will use the notation 
\beq
\label{app4daf}
\lim \int d^3 x \, j^{{\mathfrak i}0} (x) \, \varphi(y) := \int d^3 x \, j^{{\mathfrak i}0} (x) \cdot \varphi(y)
~.
\eeq
Consider the following special cases of \eqref{app4dae}:
\begin{itemize}
\item At $y=0$
\bea
\label{app4dag}
&&\int d^3 x \, G^{\mathfrak i}_{\alpha 0} \cdot \varphi(0) =  \left[ Q^{\mathfrak i}_\alpha,\varphi \right](0)~, 
\\
&&\int d^3 x \, x^{\dot\alpha \alpha} G^{\mathfrak i}_{\dot\alpha 0 }(x) \cdot \varphi(0) = - \sigma_\mu^{\dot\alpha \alpha} \sigma^\mu_{\alpha \dot \beta} \left[ \bar S^{{\mathfrak i}\dot \beta},\varphi \right] (0) = 4  \left[ \bar S^{{\mathfrak i}\dot \alpha},\varphi \right] (0) 
~.
\eea
\item At general $y\neq 0$
\bea
\label{app4dai}
&&\int d^3 x \, G^{\mathfrak i}_{\alpha 0} \cdot \varphi(y) =  \left[ Q^{\mathfrak i}_\alpha,\varphi \right](y)~, 
\\
&&\int d^3 x \, x^{\dot\alpha \alpha} G^{\mathfrak i}_{\dot\alpha 0 }(x) \cdot \varphi(y) = 
y^{\dot\alpha\alpha} \left[ Q^{\mathfrak i}_\alpha,\varphi \right](y)
+ 4  \left[ \bar S^{{\mathfrak i}\dot \alpha},\varphi \right] (y) 
~.
\eea
\item For a {\it superconformal} primary  $\left[ \bar S^{{\mathfrak i}\dot \alpha},\varphi \right] (y)=0$, hence
 \bea
\label{app4dak}
&&\int d^3 x \, G^{\mathfrak i}_{\alpha 0} \cdot \varphi(y) =  \left[ Q^{\mathfrak i}_\alpha,\varphi \right](y)~, 
\\
\label{app4daka}
&&\int d^3 x \, x^{\dot\alpha \alpha} G^{\mathfrak i}_{\alpha 0 }(x) \cdot \varphi(y) = 
y^{\dot\alpha\alpha} \left[ Q^{\mathfrak i}_\alpha,\varphi \right](y) 
~.
\eea
\end{itemize}

Now consider the operator $\left[ Q^{\mathfrak i}_\alpha,\varphi \right](y)= \left[ Q^{\mathfrak i}_\alpha,\varphi(y) \right]$ on the 3-sphere, $|y|=1$, where $\varphi$ is a superconformal primary. When we transform to $\IR \times S^3$ and work in radial quantization we perform a conformal transformation and a transformation \eqref{defsaa} to a new basis. Under this transformation 
\beq
\label{app4dal}
{\left[ Q^{\mathfrak i}_\alpha,\varphi(y) \right] }
\to \left[ \QQ^{+{\mathfrak i}}_\alpha, \widetilde{\varphi(y)} \right]
~.
\eeq
Employing \eqref{app4daka} we get
\bea
\label{app4dam}
&&y^{\dot\alpha \alpha}\left[ \QQ^{+{\mathfrak i}}_\alpha, \widetilde{\varphi(y)} \right] 
=  \left(  y^{\dot\alpha\alpha} \widetilde{ \left[ Q^{\mathfrak i}_\alpha,\varphi(y) \right]  } \right) 
= \left(\widetilde{\int d^3 x \, x^{\dot\alpha \alpha} G^{\mathfrak i}_{\alpha 0}(x) \cdot \varphi(y) }\right)
\nonumber\\
&&= \left( \widetilde{\int d^3 x \, x^{\dot\alpha \alpha} G^{\mathfrak i}_{\alpha 0}}(x) \right) \cdot \widetilde{\varphi(y)}
: = \left[ \SS^{-{\mathfrak i} \dot \alpha}, \widetilde{\varphi(y)}\right]
~.
\eea
The last equality defines the action of the supercharge $\SS^{-{\mathfrak i} \dot\alpha}$ on $\varphi(y)$, and gives the precise meaning of the expressions that we write in the main text when we express the Hamiltonian deformations (in the main text we dropped the tildes). This definition has all the properties we assumed in the main text: it acts on the external states as the `rotated' $\SS^{-{\mathfrak i}\dot\alpha}$ and satisfies the superconformal algebra relations listed in section 3 of \cite{Dolan:2002zh}. 

As a consequence of \eqref{app4dam} we get
\beq
\label{app4dan}
\varepsilon_{\dot\alpha\dot \beta} \left\{ \SS^{-{\mathfrak i} \dot\alpha}, \left[ \SS^{-{\mathfrak i} \dot \beta} , \widetilde {\varphi(y)} \right] \right\}
= \varepsilon_{\dot\alpha \dot\beta} y^{\dot \alpha \alpha} y^{\dot \beta \beta} 
\left\{ \QQ^{+{\mathfrak i}}_\alpha , \left [ \QQ^{+{\mathfrak i}}_{\beta} , \widetilde{ \varphi(y) } \right] \right \}
= \varepsilon^{\alpha \beta} \left\{ \QQ^{+{\mathfrak i}}_\alpha , \left [ \QQ^{+{\mathfrak i}}_{\beta} , \widetilde{ \varphi(y) } \right] \right \}
~.
\eeq
To get the last equality we used 
\beq
\label{app4dao}
\varepsilon_{\dot\alpha \dot\beta} y^{\dot \alpha \alpha} y^{\dot \beta \beta} 
= \varepsilon_{\dot\alpha \dot\beta} \sigma_\mu^{\dot\alpha \alpha} \sigma_\nu^{\dot \beta \beta} y^\mu y^\nu
= \varepsilon^{\alpha\beta} y^\mu y_\mu = \varepsilon^{\alpha\beta}
~.
\eeq  
At the last step we used $|y|=1$. The identity \eqref{app4dan} is the reason why we exchanged $\QQ^+$s with $\SS^-$s in the main text (as well as $\QQ^-$ with $\SS^+$ from analogous statements).

Finally, we point out that for a superconformal primary $\varphi$ with the extra property $\left[ \bar Q_{{\mathfrak i}\dot \alpha}, \varphi(y) \right] =0$ we have
\bea
\label{app4dap}
&&\int d^3 x \, \bar G_{{\mathfrak i} \dot \alpha 0} \cdot \varphi(y) =  \left[ \bar Q_{{\mathfrak i}\dot \alpha},\varphi \right](y) =0 ~, 
\\
\label{app4dapa}
&&\int d^3 x \, x^{\dot\alpha \alpha} \bar G_{{\mathfrak i} \dot\alpha 0 }(x) \cdot \varphi(y) = 
y^{\dot\alpha\alpha} \left[ \bar Q_{{\mathfrak i}\dot \alpha},\varphi \right](y)=0 
~.
\eea
The transformation of these properties to the cylinder gives 
\beq
\label{app4daq}
\left[ \SS^+_{{\mathfrak i}\dot \alpha} , \widetilde { \varphi(y) } \right] =0~, ~~
\left[ \QQ^{-\alpha}_{\mathfrak i} , \widetilde { \varphi(y) } \right] =0
~.
\eeq
These properties are also useful in the main text.

\section{Note on 1/2-BPS operators in $\NN=4$ SYM theory}
\label{half}

Ref.\ \cite{Baggio:2014ioa} showed that the curvature of 1/2-BPS chiral primary bundles in $\NN=4$ SYM theory vanishes. As we reviewed in the introduction, the proof of \cite{Baggio:2014ioa} requires the use of the 4d \tt\ equations, the non-renormalization of 3-point functions of 1/2-BPS operators and an independent proof of the tree-level identities \eqref{setaj} (see appendix C in \cite{Baggio:2014ioa} for this proof). In this appendix we present an alternative simpler proof of the vanishing curvature in the 1/2-BPS sector that does not use the above ingredients and relies only on special properties of the $\NN=4$ SYM theory.

In $\NN=4$ SYM theory there is a single complex exactly marginal coupling, the complexified Yang-Mills coupling $\tau = \frac{\theta}{2\pi} + \frac{4\pi i}{g_{YM}^2}$. On $\IR^4$ the corresponding exactly marginal interaction involves the operator
\beq
\label{halfaa}
\OO_\tau = (Q^4)^2 (Q^3)^2\cdot \varphi :=
\{ Q_-^4,[Q_+^4,\{Q_-^3,[Q_+^3,\varphi]\}]\}
~,
\eeq
where $\varphi$ is the $(0,2,0)$ 1/2-BPS highest weight superconformal primary. If $Z$ denotes one of the complex adjoint scalars of $\NN=4$ SYM theory, then $\varphi \sim \Tr[Z^2]$. We are using the notation $Q^{\mathfrak i}_\alpha$, $\bar Q_{{\mathfrak i}\dot\alpha}$ for the Poincar\'e supercharges and $S_{\mathfrak i}^\alpha$, $\bar S^{{\mathfrak i}\dot\alpha}$ for the their superconformal partners. The indices ${\mathfrak i}=1,2,3,4$ are $SU(4)_R$ indices and the $\alpha,\dot\alpha=\pm$ are spacetime spinor indices. The complex conjugate operator $\OO_{\bar \tau}$ reads
\beq
\label{halfab}
\OO_{\bar \tau} = (\bar Q_4)^2 (\bar Q_3)^2 \cdot \bar \varphi
~.
\eeq

In $\NN=4$ SYM theory we can recast the above expressions for $\OO_\tau$ and $\OO_{\bar\tau}$ as
\beq
\label{halfac}
\OO_\tau = (Q^1)^2 (Q^2)^2 \cdot \bar\varphi
~,
\eeq
\beq
\label{halfad}
\OO_{\bar \tau} = (\bar Q_1)^2 (\bar Q_2)^2 \cdot \varphi
~.
\eeq
We will now use this special property to prove that the curvature of 1/2-BPS operators vanishes.

First, let us consider the proof in the language of operators and conformal perturbation theory. In that context the expression for the curvature is \cite{Papadodimas:2009eu}
\beq
\label{halfae}
(F_{\tau\bar \tau})_{K\bar L} = \frac{1}{(2\pi)^4} \int_{|x|\leq 1}d^4 x \int_{|y|\leq 1}d^4 y
\bigg[ \langle \bar\phi_L (\infty) \OO_\tau(x) \OO_{\bar\tau}(y) \phi_K(0)\rangle
- \langle \bar\phi_L (\infty) \OO_{\bar\tau}(x) \OO_{\tau}(y) \phi_K(0)\rangle \bigg]
.
\eeq
The operators $\phi_K$ are 1/2-BPS $SU(4)_R$ highest weight superconformal primaries. Two of their defining properties are
\beq
\label{halfaea}
[Q^1_\alpha, \phi_K ]=0~, ~~
[Q^2_\alpha, \phi_K ]=0
~,
\eeq
\beq
\label{halfaeb}
[\bar Q_{3\dot\alpha}, \phi_K ]=0~,~~
[\bar Q_{4\dot\alpha}, \phi_K ]=0
~.
\eeq
The $\bar\phi_L$ are complex-conjugate lowest weight superconformal primaries, which obey the complex-conjugate versions of \eqref{halfaea}, \eqref{halfaeb}. In expression \eqref{halfae} we include implicitly a regularization prescription \cite{Ranganathan:1993vj} that excises small balls of radius $\varepsilon$ around the insertions $\phi_K$, $\bar \phi_L$. At the end of the computation we are instructed to take the limit $\varepsilon \to 0$ and discard the divergent pieces. We will argue that the expression \eqref{halfae} for $F_{\tau\bar \tau}$ vanishes at finite $\varepsilon$, hence it also vanishes after we take the limit.

Using the freedom to express $\OO_\tau$, $\OO_{\bar\tau}$ in any of the possible ways in \eqref{halfaa}-\eqref{halfad} we write
\begin{align}
\begin{split}
\label{halfaf}
&\langle \bar\phi_L (\infty) \OO_\tau(x) \OO_{\bar\tau}(y) \phi_K(0)\rangle
- \langle \bar\phi_L (\infty) \OO_{\bar\tau}(x) \OO_{\tau}(y) \phi_K(0)\rangle
\\
=&~ \langle \bar\phi_L (\infty) \left( (Q^1)^2 (Q^2)^2\cdot \bar \varphi \right)(x) \left( (\bar Q_4)^2 (\bar Q_3)^2 \cdot \bar \varphi \right)(y)  \phi_K(0)\rangle
\\
&- \langle \bar\phi_L (\infty) \left(  (\bar Q_4)^2 (\bar Q_3)^2 \cdot \bar \varphi   \right) (x) \left(   (Q^1)^2 (Q^2)^2\cdot \bar \varphi  \right) (y) \phi_K(0)\rangle
~.
\end{split}
\end{align}
We observe the following facts about the first term on the RHS of this equation. Since $\bar Q_3$ and $\bar Q_4$ annihilate $\phi_K$, \eqref{halfaeb}, we can use a superconformal Ward identity to move all the $\bar Q_3$ and $\bar Q_4$ from the $y$-insertion to the $x$-insertion. As reviewed in \cite{Papadodimas:2009eu} the $\bar \phi_L$ at infinity is hidden and there is no contribution from the $\bar Q_3$ and $\bar Q_4$ acting on it. Moreover, the $\bar Q_3$ and $\bar Q_4$ commute with the $Q^1$ and $Q^2$ supercharges already present in the $x$-insertion. Similarly, we can next move the $Q^1$ and $Q^2$ supercharges from the $x$-insertion to the $y$-insertion. As a result we find that the first term and the second term on the RHS of eq.\ \eqref{halfaf} are equal and cancel out. This proves that $F_{\tau\bar\tau}$ in \eqref{halfae} vanishes identically.

It is interesting to note that it is not possible to carry out the same manipulation in the case of 1/4-BPS states in order to prove that the corresponding curvature vanishes. Instead of \eqref{halfaea}, \eqref{halfaeb} the 1/4-BPS operators obey only
\beq
\label{halfag}
[Q^1_\alpha, \phi_K ]=0~, ~~
[\bar Q_{4\dot \alpha}, \phi_K ]=0
~.
\eeq
Hence, we cannot freely exchange the position of the $Q^2$ and $\bar Q_3$ supercharges in the expression for the curvature. This might be an indication that the curvature of 1/4-BPS operators in $\NN=4$ SYM theory does not vanish.

Finally, as a check let us show how to rederive the above result on the 1/2-BPS sector as a vanishing of the corresponding Berry curvature. We formulate the Berry curvature as 
\beq
\label{halfai}
\left( F_{\tau\bar\tau} \right)_{I\bar J} = \lim_{x\to 0} \left( \widetilde F_{\tau\bar\tau} \right)_{I\bar J}
~,
\eeq
where
\begin{align}
\begin{split}
\label{halfaj}
&\hspace{-0.1cm} \left( \widetilde F_{\tau\bar\tau} \right)_{I\bar J} = \frac{1}{(2\pi)^4} \lim_{\time_1 \to 0^-} 
\\
&\hspace{-0.1cm}\bigg[ 
\langle \bar J | \left( \QQ^{+1}\right)^2 \left( \QQ^{+2}\right)^2 \cdot \sint \bar \varphi(\time_1) \, (\HH_1 - x)^{-1} (\HH_4 - x)^{-1} \,  \left( \SS^{+}_4\right)^2 \left( \SS^{+}_3 \right)^2 \cdot \sint \bar \varphi(\time_2) |I \rangle
\\
&\hspace{-0.1cm}- \langle \bar J | \left( \SS^{+}_4\right)^2 \left( \SS^{+}_3\right)^2 \cdot \sint \bar \varphi(\time_1) \, (\HH_1 - x)^{-1} (\HH_4 - x)^{-1} \,  \left( \QQ^{+1}\right)^2 \left( \QQ^{+2}\right)^2  \cdot \sint \bar \varphi(\time_2) |I \rangle 
\bigg]
.
\end{split}
\end{align}
The definitions of the calligraphic supercharges are the same as in appendix \ref{defs}. We also defined
\beq
\label{halfaia}
\HH_1 := H - \frac{1}{2} ( 3 \RR_1 + 2 \RR_2 + \RR_3)~,~~
\HH_4 := H - \frac{1}{2} (  \RR_1 + 2 \RR_2 + 3 \RR_3)
~.
\eeq
$\RR_1$, $\RR_2$, $\RR_3$ are the Cartan generators of the $SU(4)_R$ R-symmetry. The combinations in \eqref{halfaia} are useful because the general highest weight 1/2-BPS and 1/4-BPS states $(q,p,q)$ have 
\beq
\label{halfaib}
\HH_1 |I \rangle =0~, ~~ \HH_4 |I\rangle =0
~.
\eeq
Using the properties of the 1/2-BPS external kets and bras
\bea
\label{halfaka}
&&\QQ^{+{\mathfrak i}}_\alpha |I \rangle =0~, ~~ {\mathfrak i}=1,2~, ~~~
\SS^+_{{\mathfrak i}\dot\alpha} | I \rangle=0~, ~~ {\mathfrak i}=3,4~,
\\
&&\QQ^{-\alpha}_{\mathfrak i} | I \rangle =0~, ~~ 
\SS^{-{\mathfrak i}\dot\alpha} | I \rangle =0~, ~~ {\mathfrak i}=1,2,3,4
~,
\eea
\bea
\label{halfakb}
&& \langle \bar J | \QQ^{-\alpha}_{\mathfrak i} =0~, ~~ {\mathfrak i}=1,2~, ~~~
\langle \bar J | \SS^{-{\mathfrak i}\dot\alpha} =0~, ~~ {\mathfrak i}=3,4~,
\\
&&\langle \bar J |  \QQ^{+{\mathfrak i}}_\alpha =0~, ~~ 
\langle \bar J |  \SS^{+{\mathfrak i}\dot\alpha}=0~, ~~ {\mathfrak i}=1,2,3,4
\eea
and the superconformal algebra relations
\beq
\label{halfala}
[\HH_1, \SS^+_3]=[\HH_1, \SS^+_4]=0~, ~~
[\HH_4 ,\SS^+_3]= \SS_3^+~, ~~
[\HH_4, \SS^+_4]= - \SS_4^+
~,
\eeq
\beq
\label{halfalb}
[\HH_4, \QQ^{+1}]=[\HH_4, \QQ^{+2}]=0~, ~~
[\HH_1 ,\QQ^{+1}]= - \QQ^{+1}~, ~~
[\HH_1, \QQ^{+2}]=  \QQ^{+2}
~,
\eeq
\beq
\label{halfalc}
[\QQ^{+{\mathfrak i}},\SS^+_{\mathfrak j}]=0~{\rm for}~ {\mathfrak i}=1,2~ {\rm and}~ {\mathfrak j}=3,4
\eeq
we can move on the first term of the RHS of \eqref{halfaj} the $\SS^+_4$, $\SS^+_3$ supercharges from the $\time_2$ to the $\time_1$ insertion and the $\QQ^{+1}$, $\QQ^{+2}$ supercharges from the $\time_1$ insertion to the $\time_2$ insertion. That shows that the first term on the RHS of \eqref{halfaj} equals the second term and therefore we recover (much in the same way as in the operator formalism) that the curvature vanishes.

As a check of the statements in appendix \ref{HamiltonDefs} we note that we can also obtain this result by writing
\begin{align}
\begin{split}
\label{halfam}
&\hspace{-0.1cm}\left( \widetilde F_{\tau\bar\tau} \right)_{I\bar J} = \frac{1}{(2\pi)^4} \lim_{\time_1 \to 0^-} 
\\
&\hspace{-0.1cm}\bigg[ 
\langle \bar J | \left( \SS^{-3}\right)^2 \left( \SS^{-4}\right)^2 \cdot \sint \varphi(\time_1) \, (\HH_1 - x)^{-1} (\HH_4 - x)^{-1} \,  \left( \QQ^{-}_1\right)^2 \left( \QQ^{-}_2 \right)^2 \cdot \sint \varphi(\time_2) |I \rangle
\\
&\hspace{-0.1cm}- \langle \bar J | \left( \QQ^{-}_1\right)^2 \left( \QQ^{-}_2\right)^2 \cdot \sint \varphi(\time_1) \, (\HH_1 - x)^{-1} (\HH_4 - x)^{-1} \,  \left( \SS^{-3}\right)^2 \left( \SS^{-4}\right)^2  \cdot \sint \varphi(\time_2) |I \rangle 
\bigg].
\end{split}
\end{align}
In that case we must use the superconformal algebra relations
\beq
\label{halfana}
[\HH_1, \SS^{-3}]=[\HH_1, \SS^{-4}]=0~, ~~
[\HH_4 ,\SS^{-3}]= - \SS^{-3}~, ~~
[\HH_4, \SS^{-4}]=  \SS^{-4}
~,
\eeq
\beq
\label{halfanb}
[\HH_4, \QQ^-_{1}]=[\HH_4, \QQ^-_{2}]=0~, ~~
[\HH_1 ,\QQ^-_{1}]=  \QQ^-_{1}~, ~~
[\HH_1, \QQ^-_{2}]= - \QQ^-_{2}
~,
\eeq
\beq
\label{halfanc}
[\QQ^-_{{\mathfrak i}},\SS^{-{\mathfrak j}}]=0~{\rm for}~ {\mathfrak i}=1,2~ {\rm and}~ {\mathfrak j}=3,4
~.
\eeq

\end{appendix}

\newpage
\providecommand{\href}[2]{#2}

\end{document}